\documentclass[12pt,draftclsnofoot,onecolumn]{IEEEtran}
\IEEEoverridecommandlockouts
\usepackage{cite}
\usepackage[tbtags]{amsmath}
\usepackage{amssymb,amsfonts}
\allowdisplaybreaks[4]
\usepackage{algorithmic}
\usepackage{graphicx}
\usepackage{textcomp}
\usepackage{xcolor}
\usepackage{booktabs}
\usepackage{pgfplots}
\usepackage{float}
\usepackage{stfloats}
\usepackage{tikz}
\usepackage{makecell}
\usepackage[linesnumbered,ruled]{algorithm2e}
\usepackage[thmmarks,amsmath]{ntheorem} 
\usepackage{makecell}
\usepackage{multirow}
\usepackage{verbatim}
\usepackage{subfigure}
\usepackage{lettrine}
\usepackage{physics}

\usepackage{threeparttable}
\usepackage{booktabs}
\usepackage{mathtools} 

\newtheorem{theorem}{Theorem}
\newtheorem{lemma}{Lemma}
\newtheorem{corollary}{Corollary}

\newcommand*{\circled}[1]{\lower.7ex\hbox{\tikz\draw (0pt, 0pt)%
    circle (.5em) node {\makebox[1em][c]{\small #1}};}}

 \qedsymbol{\ensuremath{\square}} 
{   
   \theoremstyle{nonumberplain}
   \theoremheaderfont{\bfseries}
   \theorembodyfont{\normalfont}
   \theoremsymbol{\ensuremath{\square}} 
   
}

\def\BibTeX{{\rm B\kern-.05em{\sc i\kern-.025em b}\kern-.08em
    T\kern-.1667em\lower.7ex\hbox{E}\kern-.125emX}}
\begin{document}

\title{\huge  Non-asymptotic Sphere-Packing Bounds for Finite-Blocklength Scalar Gaussian Multiple-Access and Broadcast Channels}

\author{\normalsize 
Kaixuan Bao, \textit{Graduate Student Member, IEEE}, Wei Xu, \textit{Fellow, IEEE},\\and Xiaohu You, \textit{Fellow, IEEE}

\thanks{

Kaixuan Bao, Wei Xu, and Xiaohu You are with the National Mobile Communications Research Laboratory (NCRL), Southeast University, Nanjing 210096, China, and are also with Purple Mountain Laboratories, Nanjing 211111, China (e-mail: {kxbao, wxu, xhyu}@seu.edu.cn).
}
}

\maketitle

\begin{abstract}
This paper investigates non-asymptotic achievable rate regions of Gaussian multiple-access channels (MACs) and broadcast channels (BCs) in the finite-blocklength regime. By employing a proposed volume-uniform random coding (VURC) codebook that distributes codewords uniformly within an $n$-dimensional power sphere, together with successive interference cancellation (SIC) decoding, we derive a closed-form achievability bound, a scheme-specific lower bound on the average error probability, and a fourth-order asymptotic characterization. The proposed SIC achievability bound admits a simple and analytically interpretable closed-form expression, enabling direct performance evaluation without iterative computation. The structure of the proposed SIC asymptotic bound offers a highly extensible framework, allowing advances in single-user finite-blocklength theory to be systematically incorporated into the multiuser setting. Futhermore, to characterize the performance benefit of joint decoding, we further derive numerical and closed-form achievability bounds for the two-user Gaussian MAC under joint nearest-neighbor (JNN) decoding. Numerical results illustrate the performance differences among the proposed SIC and JNN bounds and existing benchmarks.
\end{abstract}

\begin{IEEEkeywords}
Finite blocklength, sphere packing, achievability, channel coding, multiple access channel (MAC), broadcast channel (BC).
\end{IEEEkeywords}

\section{Introduction}

The envisioned sixth generation (6G) of wireless communication networks are expected to impose two significant requirements: ultra-high connection density and ultra-low communication latency~\cite{you6G}. Both stringent demands motivate intensive research on multiuser scenarios under the finite blocklength regime, where the classical asymptotic assumption of infinite codeword length no longer holds for the desired vanishing error probability~\cite{10147111}. In this regime, understanding the relationship between the maximum achievable rate of the channel, i.e., $R(\gamma,n,\epsilon)$, given a signal-to-noise ratio (SNR) $\gamma$, a blocklength $n$, and an average error probability $\epsilon$, becomes crucial for designing practical coding and decoding schemes that can meet the expected 6G requirements~\cite{10024766}.

In this paper, we consider a multiuser communication scenario where a single-antenna base station (BS) communicates with $K$ users, each equipped with a single antenna. In the downlink transmission, the BS acts as a single transmitter serving $K$ receivers, which constitutes a broadcast channel (BC). Conversely, in the uplink, $K$ users transmit simultaneously to the BS, forming a multiple access channel (MAC).

When $K=1$, both channels reduce to the single-input single-output (SISO) case. Finite blocklength studies for this case fall into two regimes: (1) the central limit theorem (CLT) regime, denoted by $R_\gamma(n,\epsilon)$, which fixes $n$ and $\epsilon$ and then derives $R$ that converges to the channel capacity $C$ as $n$ tends to infinity, and (2) the large deviation (LD) regime, denoted by $\epsilon_\gamma(n,R)$, which fixes $n$ and $R$ and then analyzes $\epsilon$ that vanishes exponentially in $n$~\cite{8187464}. In the CLT regime, Polyanskiy, Poor, and Verd\'u~\cite[Th.~54]{PPV}showed in 2010 that
\begin{align}
    R_\gamma(n,\epsilon)
    =
    C(\gamma)
    -
    \sqrt{\frac{V(\gamma)}{n}}
    Q^{-1}(\epsilon)
    +
    o\!\left(n^{-1/2}\right),
\end{align}
where
\begin{align}
    C(\gamma)
    &=
    \frac{1}{2}\log(1+\gamma),\\
    V(\gamma)
    &=
    \frac{\gamma(\gamma+2)}
    {2(\gamma+1)^2}\log_2^2 e,
\end{align}
and $Q(\cdot)$ denotes the Gaussian $Q$-function. The second term on the right-hand side is termed the second-order result. Subsequently, the third-order result was characterized by Tan and Tomamichel~\cite[Th.~1]{Tan}, while a more refined higher-order result was given by Yavas, Kostina, and Effros~\cite[Th.~5]{YKE}. In the LD regime, Shannon's sphere-packing result (SP59)~\cite{SP59} was one of the earliest results and Gallager subsequently developed refined exponential bounds~\cite{SP67,1967}. Different from the CLT regime, which characterizes the rate backoff from capacity for a prescribed error probability, the LD regime directly describes the exponential dependence of the error probability on the blocklength in closed form. This viewpoint is particularly relevant to the ultra-short blocklength regime considered in this work, where both the error exponent and its finite corrections remain significant.

Building upon these fundamental single-user results, it is natural to seek finite blocklength characterizations for multiuser channels. However, compared with the single-user case, analyzing the performance limits of the MAC and BC under finite blocklength constraints presents substantial new challenges, as the overall error event cannot be directly factorized. For the Gaussian MAC, MolavianJazi and Laneman~\cite{ML} derived achievability bounds for the two-user case based on the dependency-testing (DT) bound. Scarlett \emph{et al.}~\cite{scarlettMAC} studied more general multiuser settings using maximum-likelihood (ML) decoding, while Yavas, Kostina, and Effros~\cite{YKEMAC} derived results based on the MAC random-coding union (RCU) bound and established a refined third-order characterization. Wang \emph{et al.}~\cite{zhoulin} further generalized the second-order result to a mismatched non-Gaussian MAC. For the Gaussian BC, second-order achievable and converse rate regions based on superposition and rate splitting were investigated in~\cite{Tuninetti2023}. Recent work by {\"U}nsal and Gorce~\cite{UnsalGorce2026} established finite-blocklength achievability bounds for dirty-paper coding using dependence-testing and $\kappa\beta$ techniques. These existing results provide important and information-theoretic characterizations by retaining the coupled multiuser error events. However, their finite-$n$ evaluation typically involves multidimensional probability or expectation calculations, and the analytical complexity increases rapidly with the number of users. Table~\ref{table of summary} provides a comparison among related work in this area, and with the result to be presented in this paper.

\begin{table}
    \centering
    \caption{Comparison with Related Works}
\begin{threeparttable}
    \begin{tabular}{|c|c|c|c|c|c|}\hline
          Year&Paper&Bound$^1$&Channel$^2$&Number of users&\textbf{Order}\\\hline
 2015& MolavianJazi and Laneman \cite{ML}
 & Ach.+Asymp.& AWGN MAC &$2$&\textbf{2}\\\hline
 
 2015& Scarlett \textit{et al.} \cite{scarlettMAC}
 & Asymp.& DMC/AWGN MAC&$\geq 2$&\textbf{2}\\\hline
 
 2016& Fong and Tan \cite{Tanconv}
 & Conv.& AWGN MAC&$\geq 2$&\textbf{2}\\\hline
 
 2021& Yavas, Kostina, and Effros \cite{YKEMAC}
 & Ach.+Asymp.& AWGN MAC&$\geq 2$&\textbf{3}\\\hline
 
 2022& Kosut \cite{Kosutconv}
 & Conv.& DMC/AWGN MAC&$2$&\textbf{2}\\\hline
 
 2023& Tuninetti \textit{et al.} \cite{Tuninetti2023}
 & Ach.+Conv.+Asymp.& AWGN BC &$2$&\textbf{2}\\\hline
 
 2025& Wang \textit{et al.} \cite{zhoulin}
 & Asymp.& mismatched MAC&$\geq 2$&\textbf{2}\\\hline
 
 2026& {\"U}nsal and Gorce \cite{UnsalGorce2026}
 & Ach.& AWGN BC &$2$& 2\\\hline
 
 2027& This paper
 & Ach.+Asymp.& AWGN MAC/BC&$\geq 2$&\textbf{4}\\\hline
    \end{tabular}
    \begin{tablenotes}
        \footnotesize
        \item[1] `Ach.' refers to achievability, `Conv.' refers to converse, and `Asymp.' refers to asymptotic.
        \item[2] AWGN refers to an additive white Gaussian noise channel, DMC refers to a discrete memoryless channel, MAC refers to a multiple-access channel, and BC refers to a broadcast channel.
    \end{tablenotes}
    \label{table of summary}
\end{threeparttable}
\end{table}

Although rigorous finite-blocklength characterizations for multiuser channels have been developed, several issues remain worth further investigation. First, existing Gaussian MAC analyses predominantly employ spherical-shell codebooks, where all codewords are restricted to the boundary of the power sphere. The additional radial degree of freedom available inside the power sphere, and its potential role in multiuser codebook construction under finite blocklengths, have not been characterized. Second, existing bounds typically retain the coupled multiuser error events and are consequently expressed in terms of multidimensional probability. While such formulations provide important theoretic benchmarks, they generally do not admit simple closed-form expressions, and their analytical complexity increase rapidly with the number of users. Third, existing multiuser asymptotic studies mainly focus on second- and third-order characterizations, whose derivations rely on the joint statistical structure of multiple information-density terms. As a result, extending a refined single-user asymptotic result to a general multiuser setting usually requires a new multi-dimensional analysis. These observations motivate further investigation of finite-blocklength Gaussian MACs and BCs from the perspectives of codebook geometry, explicit non-asymptotic closed-form characterization, and extensible higher-order analysis.

\subsection{Contributions}

In this paper, we develop a volume-uniform random coding (VURC)-based finite-blocklength framework for the Gaussian MAC and BC. Compared with existing studies, the main contributions can be summarized in the following three aspects.

\emph{1) Multiuser VURC codebook design and its geometric advantage:}
We propose a multiuser codebook construction based on VURC and superposition coding under the maximum power constraint. Different from the conventional spherical-shell codebook, VURC exploits
the entire interior of the power sphere. By constructing multiple radially separated packing layers, we rigorously show that the sphere-constrained codebook possesses a strictly larger finite-dimensional packing
cardinality than its spherical-shell counterpart under the same minimum distance requirement. This result establishes a theoretical geometric advantage of VURC for multiuser finite-blocklength transmission.

\emph{2) Closed-form non-asymptotic achievable bounds with SIC and JNN:}
Based on the proposed VURC codebook, we first derive an explicit closed-form non-asymptotic achievability bound for the Gaussian MAC and BC with an arbitrary number of users under SIC decoding. The main technical step is an SOSD-based Gaussian replacement of the non-Gaussian aggregate interference-plus-noise term, which reduces each SIC stage to an equivalent single-user problem and yields a simple analytical expression with low computational complexity. To address the suboptimal issue of SIC relative to joint decoding, we further derive a JNN characterization for the two-user Gaussian MAC, including closed-form and numerical bounds. For the considered VURC+SIC setup, we also derive a lower bound on the average error probability.

\emph{3) Extensible higher-order asymptotic characterization:}
Based on the equivalent-SINR structure revealed by the non-asymptotic analysis, we establish a unified framework that systematically maps single-user finite-blocklength asymptotic results to the considered Gaussian MAC and BC. The second-order term captures the dominant dispersion penalty, whereas the third- and fourth-order terms refine the residual finite-blocklength corrections and provide more accurate performance descriptions at short blocklengths. More importantly, the resulting structure separates the multiuser
interference effect, represented by the equivalent SINR, from the single-user finite-blocklength correction, thereby allowing future advances in single-user asymptotic analysis to be directly incorporated into the multiuser framework.

The remainder of the paper is organized as follows. Section~II presents the system model for the Gaussian MAC and BC, along with the associated encoder and decoder designs. Section~III introduces the proposed multiuser VURC codebook and its geometric properties. Section~\uppercase\expandafter{\romannumeral4} introduces the main results, including two bounds under SIC, two bounds under JNN, and a series of extensible asymptotic bounds. Numerical evaluations are included in Section~\uppercase\expandafter{\romannumeral5}. Section~\uppercase\expandafter{\romannumeral6} concludes this paper.

\subsection{Notation}
In this paper, deterministic scalars and vectors are denoted by italic lowercase letters and boldface lowercase letters, respectively, e.g., $a$ and $\mathbf a$. Random variables and random vectors are denoted by italic uppercase letters and boldface uppercase letters, respectively, e.g., $A$ and $\mathbf A$. The Euclidean norm of a vector is denoted by $\|\cdot\|$, with $\|\cdot\|_2$ used equivalently when the norm needs to
be emphasized. The symbols $\mathbf 0$ and $\mathbf I_n$ denote, respectively, the all-zero vector of the appropriate dimension and the $n\times n$ identity matrix. We define
$\mathbb R_+\triangleq[0,\infty)$, and vector inequalities are understood componentwise unless otherwise specified.

The expectation and probability operators are denoted by $\mathbb E\{\cdot\}$ and $\Pr\{\cdot\}$, respectively. A subscript on either operator specifies the random variable or distribution with respect to which the operation is taken. The indicator function of an
event $\mathcal E$ is denoted by $\mathbf 1\{\mathcal E\}$. The notation $X\sim\mathcal N(\mu,\sigma^2)$ denotes a Gaussian random variable with mean $\mu$ and variance
$\sigma^2$, with the corresponding vector notation $\mathbf X\sim\mathcal N(\boldsymbol\mu,\mathbf\Sigma)$. Moreover, $\mathbf X\sim\mathrm{Unif}[\mathcal A]$ means that $\mathbf X$ is uniformly distributed over the set $\mathcal A$.

The probability density function (PDF) and cumulative distribution function (CDF) of a random variable $X$ are denoted by $f_X(x)$ and $F_X(x)$, respectively, and we further define $S_X(x)\triangleq\int_{-\infty}^{x}F_X(t)\,\dd t.$ For two nonnegative random variables $X$ and $Y$, we write $X\preceq_{\rm icx}Y$ if $\mathbb E\{g(X)\}\leq\mathbb E\{g(Y)\}$ for every non-decreasing convex function $g$ for which the expectations exist. For two random vectors $\mathbf X$ and $\mathbf Y$ of the same dimension, we write
$\mathbf X\preceq_{\rm cx}\mathbf Y$ if $\mathbb E\{\phi(\mathbf X)\}\leq\mathbb E\{\phi(\mathbf Y)\}$ for every convex function $\phi$ for which the expectations exist.

The notation $[M]$ represents the set $\{1,2,\ldots,M\}$, and $|\mathcal A|$ denotes the cardinality of a finite set $\mathcal A$. $\Gamma(z)=\int_0^{\infty} t^{z-1}\mathrm{e}^{-t}\,\mathrm{d} t$ denotes the gamma function, and $\Gamma(z,a)=\int_a^{\infty} t^{z-1}\mathrm{e}^{-t}\,\mathrm{d} t$ and $\gamma(z,a)=\int_0^{a} t^{z-1}\mathrm{e}^{-t}\,\mathrm{d} t$ are respectively the upper and lower incomplete gamma functions. Similarly, $B(x,y)=\int_0^1 t^{x-1}(1-t)^{y-1}\,\text{d}t$ denotes the beta function and $I_z(x,y)=B^{-1}(x,y)\int_0^z t^{x-1}(1-t)^{y-1}\,\text{d}t$ is the normalized incomplete beta function.

We use the big-$O$ notation $f(n)=O(g(n))$ if and only if there exist constants $c>0$ and $n_0$ such that $|f(n)|\leq c|g(n)|$ for all $n>n_0$, and the little-$o$ notation $f(n)=o(g(n))$ if and only if for every $c>0$ there exists a constant $n_0$ such that $|f(n)|\leq c|g(n)|$ for all $n>n_0$.

\section{System Models}
\label{sec:system_model}

\begin{figure*}
\centering
\subfigure[The broadcast channel]{
\includegraphics[width=7cm]{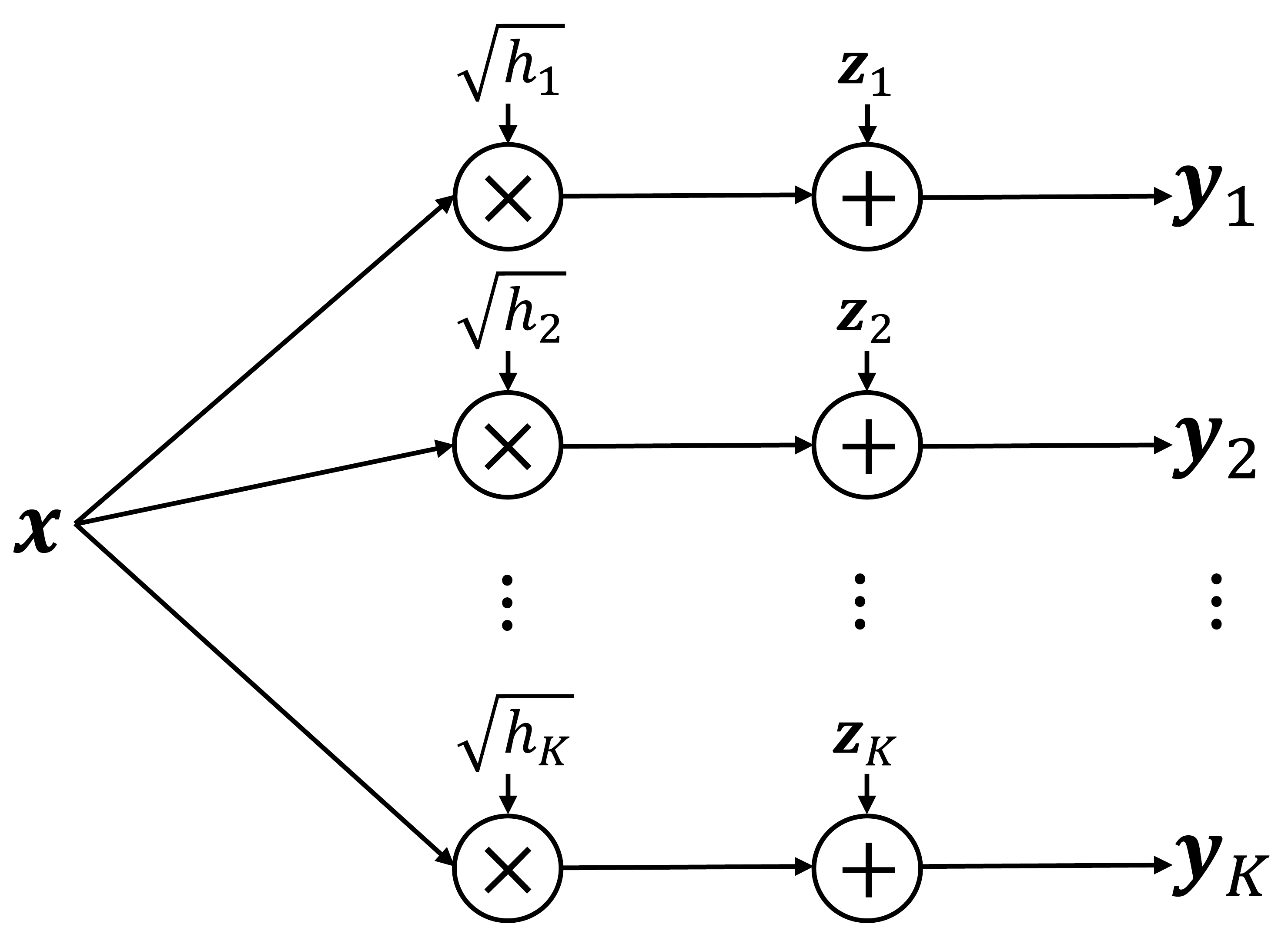}
}
\subfigure[The multiple access channel]{
\includegraphics[width=6.8cm]{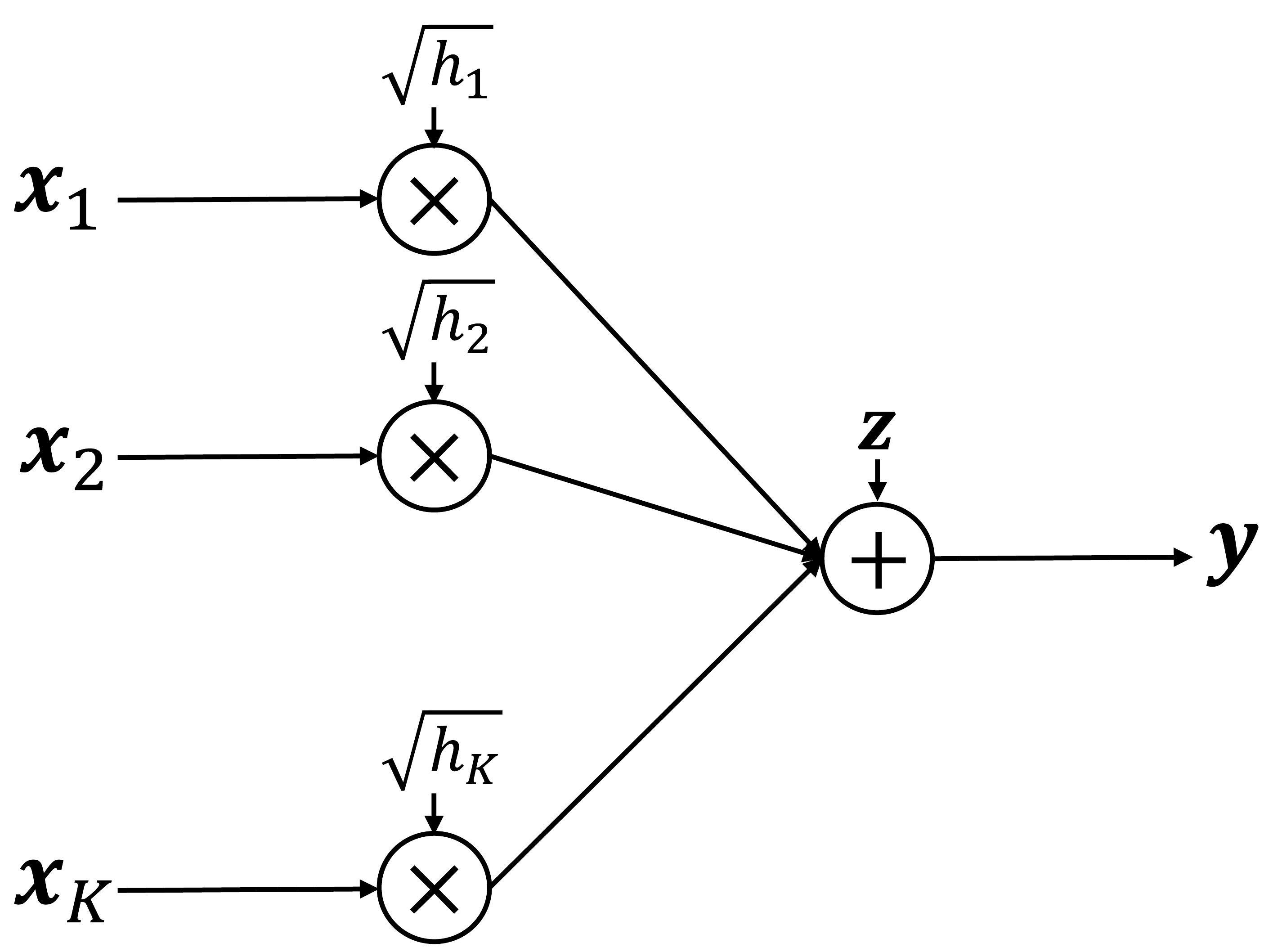}
}
\caption{System models.}
\label{system models}
\end{figure*}

We consider two different systems as shown in Fig.~\ref{system models}. The left one, i.e., Fig.~\ref{system models}(a), corresponds to a BC, where one transmitter with one antenna sends a codeword to $K$ receivers simultaneously. Each receiver has one antenna. The codeword is denoted by $\mathbf{x}\in\mathbb{R}^{n}$, where $n$ is the blocklength. The transmitted signal passes a flat fading AWGN channel, where the channel power gain is $h_i,i=1,2,\cdots,K$\footnote{The channel gain is induced by Rayleigh fading. In addition, this paper focuses on the finite blocklength regime where $n$ is small. Therefore, the channel gain for one transmission, i.e., $n$ channel uses, is assumed to be a constant.} and the additive Gaussian noise is $\mathbf{z}_i\in\mathbb{R}^{n}$. The right one, i.e., Fig.~\ref{system models}(b), corresponds to a MAC, where $K$ users each with one antenna send $K$ codewords to one receiver simultaneously; the codeword of the $i^{\rm th}$ transmitter is denoted by $\mathbf{x}_i\in\mathbb{R}^{n},i=1,2,\cdots,K$. All transmitted signals pass the AWGN channel, where the channel gain for the $i^{\rm th}$ user is denoted by $h_i,i=1,2,\cdots,K$ and the additive Gaussian noise at the MAC receiver is $\mathbf{z}\in\mathbb{R}^{n}$. Thus, the two system models are described as
\begin{alignat}{2}
    \text{BC:}&\quad \mathbf{y}_i &&=\sqrt{h_i}\mathbf{x}+\mathbf{z}_i,\label{BCd}\\
    \text{MAC:}& \quad\mathbf{y}  &&=\sum_{i=1}^{K}\sqrt{h_i}\mathbf{x}_i+\mathbf{z}.\label{MACd}
\end{alignat}
The noise power of receiver $i$ in the BC is denoted by $\sigma_i^2$, while the noise power of the single receiver in the MAC is denoted by $\sigma^{2}$. We assume that both the transmitter(s) and the receiver(s) have perfect channel state information (CSI), i.e, perfect CSI at the transmitter (CSIT) and perfect CSI at the receiver (CSIR).

In the following, we describe the encoding and decoding process. Without loss of generality, suppose that $h_1\geq h_2\geq\cdots\geq h_K$, which means the channel gains decrease along the user index. 

\subsection{BC}

\subsubsection{Codebook}

For user $i$, let $\mathcal C_i=\left\{\mathbf{x}_i(1),\ldots,\mathbf{x}_i(M_i)\right\},M_i=2^{nR_i}$ denote its constituent codebook. We employ the VURC construction introduced in \cite{TIT}. Specifically, the codewords of $\mathcal C_i$ are independently generated according to the uniform distribution over the $n$-dimensional hypersphere
\begin{align}
    \mathcal B_n(\sqrt{P_i})
    =
    \left\{
        \mathbf{x}\in\mathbb R^n:
        \|\mathbf{x}\|^2\leq P_i
    \right\},
\end{align}
and therefore every constituent codeword satisfies the layerwise maximum-power constraint
\begin{align}
    \|\mathbf{x}_i(m_i)\|^2
    \leq
    P_i,
    \qquad
    \forall m_i\in[M_i].
    \label{eq:BC-layer-power}
\end{align}
Accordingly, the nominal SNR associated with the $i$th layer at receiver $i$ is defined as
\begin{align}
    \gamma_{i,\mathrm{BC}}
    \triangleq
    \frac{h_iP_i}{\sigma_i^2}.
\end{align}
A more detailed description of the proposed multiuser VURC codebook is presented in Section~III.

\subsubsection{Encoder}
For user $i$, the encoder is defined as
\begin{align}
    \mathcal E_i:
    [M_i]
    \rightarrow
    \mathcal C_i,
    \qquad
    \mathcal E_i(m_i)
    =
    \mathbf{x}_i(m_i).
\end{align}
The transmitter employs superposition coding, and the transmitted block is
\begin{align}
    \mathbf{x}
    =
    \sum_{i=1}^{K}
    \mathbf{x}_i(m_i).
    \label{eq:BC-superposition}
\end{align}

\subsubsection{Decoder}
We employ SIC with minimum-distance decoding. Under the decoding order $1,\ldots,K$, receiver $i$ successively decodes the constituent codewords for users $1,\ldots,i-1$ before decoding its own codeword. After removing the previously decoded layers, the signal used for decoding user $i$ is
\begin{align}
    \tilde{\mathbf{y}}_i
    =
    \mathbf{y}_i
    -
    \sqrt{h_i}
    \sum_{j=1}^{i-1}
    \hat{\mathbf{x}}_{j,i},
    \label{eq:BC-SIC-signal}
\end{align}
where $\hat{\mathbf{x}}_{j,i}$ denotes the estimate of the $j$th constituent codeword at receiver $i$. The remaining layers
$i+1,\ldots,K$ are treated as interference. Then, the desired codeword is decoded according to
\begin{align}
    \hat{m}_i
    =
    \arg\min_{m_i\in[M_i]}
    \left\|
        \tilde{\mathbf{y}}_i
        -
        \sqrt{h_i}\mathbf{x}_i(m_i)
    \right\|^2.
    \label{eq:BC-min-distance}
\end{align}

\subsection{MAC}

\subsubsection{Codebook}
For the MAC, each transmitter independently employs a VURC codebook $\mathcal C_i$ with $M_i=2^{nR_i}$ codewords generated uniformly over $\mathcal B_n(\sqrt{P_i})$. Hence,
\begin{align}
    \|\mathbf{x}_i(m_i)\|^2
    \leq
    P_i,
    \qquad
    \forall m_i\in[M_i],
    \label{eq:MAC-user-power}
\end{align}
is satisfied independently by each transmitter. Accordingly, the corresponding nominal SNR is
\begin{align}
    \gamma_{i,\mathrm{MAC}}
    \triangleq
    \frac{h_iP_i}{\sigma^2}.
\end{align}
The detailed VURC construction is again deferred to Section~III.

\subsubsection{Encoder}
Each user independently maps its message $m_i$ to $\mathbf{x}_i(m_i)\in\mathcal C_i$ and transmits the resulting codeword. Unlike the BC, no signal superposition is performed at a common transmitter; superposition occurs naturally over the MAC as in \eqref{MACd}.

\subsubsection{SIC Decoder}
For the SIC decoder, without loss of generality, we consider the decoding order $1,\ldots,K$. After successfully decoding the first $j-1$ users, the receiver forms
\begin{align}
    \tilde{\mathbf{y}}_j
    =
    \mathbf{y}
    -
    \sum_{\ell=1}^{j-1}
    \sqrt{h_\ell}
    \hat{\mathbf{x}}_\ell.
    \label{eq:MAC-SIC-signal}
\end{align}
The signals of users $j+1,\ldots,K$ are treated as interference, and user $j$ is decoded according to
\begin{align}
    \hat{m}_j
    =
    \arg\min_{m_j\in[M_j]}
    \left\|
        \tilde{\mathbf{y}}_j
        -
        \sqrt{h_j}\mathbf{x}_j(m_j)
    \right\|^2.
    \label{eq:MAC-SIC-decoder}
\end{align}

\subsubsection{JNN Decoder}
In addition to SIC, we also consider JNN decoding for the MAC as JNN is strictly better than SIC in term of performance. Rather than decoding the users sequentially, the JNN decoder jointly searches over all message tuples:
\begin{align}
    \left(
        \hat{m}_1,\ldots,\hat{m}_K
    \right)
    =
    \arg\min_{
        (m_1,\ldots,m_K)
        \in
        [M_1]\times\cdots\times[M_K]
    }
    \left\|
        \mathbf{y}
        -
        \sum_{i=1}^{K}
        \sqrt{h_i}\mathbf{x}_i(m_i)
    \right\|^2.
    \label{eq:MAC-JNN-decoder}
\end{align}
For the Gaussian MAC with equiprobable messages, this Euclidean JNN rule coincides with maximum-likelihood decoding for the prescribed codebooks.

With regard to SIC decoder, for a fixed collection of codebooks $\mathcal C\triangleq(\mathcal C_1,\ldots,\mathcal C_K)$, the decoding error probability of user $i$ is defined as
\begin{align}
    \epsilon_i(\mathcal C)
    \triangleq
    \Pr\left\{
        \hat m_i\neq m_i
        \,\middle|\,
        \mathcal C
    \right\},
    \label{eq:fixed-codebook-error}
\end{align}
where the message indices $m_j$ are independently and uniformly distributed over $[M_j]$, and the probability is taken over the transmitted message tuple and the channel noise. With regard to JNN decoder, the joint decoding error probability is defined as
\begin{align}
    \epsilon_{\mathrm J}(\mathcal C)
    \triangleq
    \Pr\left\{
        (\hat m_1,\cdots,\hat m_K)
        \neq
        (m_1,\cdots,m_K)
        \,\middle|\,
        \mathcal C
    \right\},
    \label{eq:jnn-fixed-codebook-error}
\end{align}
where the probability is taken over the uniformly distributed message pair and the channel noise. 

\section{A Multiuser VURC Codebook Design}
\label{sec:multiuser_vurc}

In the previous section, we introduced the channel models and the message transmission frameworks for the Gaussian MAC and BC. In this section, we separately investigate the codebook design. In particular, we extend the VURC codebook introduced for the single-user additive-noise channel in \cite{TIT} to the multiuser setting. We first define the constituent VURC codebooks for the MAC and BC and discuss their power constraints. We then establish a geometric advantage of VURC over conventional spherical-shell codebooks. Finally, we show that the multiuser setting enlarges the rate region in which the additional radial degree of freedom offered by VURC can be exploited.

\subsection{Multiuser VURC Codebook Design and Power Constraints}
\label{subsec:multiuser_vurc_design}

Consider a $K$-user system operating over either a MAC or a BC. For the $j$th user or constituent layer, let $\mathcal C_j=\left\{\mathbf x_j(1),\ldots,\mathbf x_j(M_j)\right\}$ and $M_j=2^{nR_j}$, where $n$ is the blocklength, $R_j$ is the coding rate, and $M_j$ is the corresponding codebook size. We define the $n$-dimensional hypersphere and its boundary as
\begin{align}
    \text{VURC:}\,\mathcal B_n(A_j)
    &\triangleq
    \left\{
        \mathbf x\in\mathbb R^n:
        \|\mathbf x\|\leq A_j
    \right\},\\
    \text{Spherical shell:}\,\mathcal S_n(A_j)
    &\triangleq
    \left\{
        \mathbf x\in\mathbb R^n:
        \|\mathbf x\|=A_j
    \right\},
    \label{eq:vurc-ball-sphere}
\end{align}
where $A_j>0$ denotes the maximum radius associated with the $j$th constituent codebook.

Following the single-user VURC construction in \cite{TIT}, the codewords of $\mathcal C_j$ are independently generated according to the uniform distribution over $\mathcal B_n(A_j)$, i.e.,
\begin{align}
    f_{\mathbf X_j}(\mathbf x)
    =
    \frac{
        \mathbf 1\{\|\mathbf x\|\leq A_j\}
    }{
        V_n(A_j)
    },
    \label{eq:vurc-vector-pdf}
\end{align}
where
\begin{align}
    V_n(A)
    =
    \frac{
        \pi^{n/2}
    }{
        \Gamma\left(\frac n2+1\right)
    }
    A^n,
    \label{eq:vurc-ball-volume}
\end{align}
is the volume of an $n$-dimensional hypersphere of radius $A$. Denote the magnitude of a VURC codeword by $T_j\triangleq\|\mathbf X_j\|$. Its PDF is
\begin{align}
    f_{T_j}(t)
    =
    \frac{
        nt^{n-1}
    }{
        A_j^n
    },
    \qquad
    0\leq t\leq A_j.
    \label{eq:vurc-radius-pdf}
\end{align}
Accordingly,
\begin{align}
    \mathbb E\left[
        \|\mathbf X_j\|^2
    \right]
    &=
    \int_0^{A_j}
    t^2 f_{T_j}(t)\,\mathrm dt\\
    &=
    \frac{n}{n+2}A_j^2.
    \label{eq:vurc-average-power}
\end{align}

For comparison, a conventional spherical-shell random codebook with the same radius $A_j$ generates every codeword on $\mathcal S_n(A_j)$ and therefore satisfies $ \|\mathbf X_j^{\mathrm{sh}}\|^2=A_j^2$ for every codeword realization. Consequently, the maximum and average power constraints coincide for spherical-shell codebooks, whereas they generally correspond to different radii for the VURC construction. In the following, we briefly distinguish the maximum and ensemble-average power constraints for VURC.

\subsubsection{Maximum Power Constraint}

Suppose that the $j$th constituent codebook is subject to the maximum power constraint
\begin{align}
    \|\mathbf x_j(m)\|^2
    \leq
    P_{j},
    \qquad
    \forall m\in[M_j].
    \label{eq:vurc-max-constraint}
\end{align}
The corresponding VURC radius is $A_j^2=P_{j}$. It follows immediately from \eqref{eq:vurc-average-power} that
\begin{align}
    \mathbb E\left[
        \|\mathbf X_j\|^2
    \right]
    =
    \frac{n}{n+2}P_{j}.
    \label{eq:vurc-max-average}
\end{align}

\subsubsection{Average Power Constraint}

For completeness, we also consider the ensemble-average power constraint
\begin{align}
    \mathbb E\left[
        \|\mathbf X_j\|^2
    \right]
    \leq
    P_j,
    \label{eq:vurc-avg-constraint}
\end{align}
where the expectation is taken over the random VURC ensemble and the uniformly distributed transmitted message. From \eqref{eq:vurc-average-power}, the full average power budget is utilized by choosing $A_j^2=\frac{n+2}{n}P_j$. In contrast, a spherical-shell codebook under the same average power constraint has radius $A_{j,\mathrm{sh}}^2=P_j$. Hence, we have $A_{j,\mathrm{VURC}}>A_{j,\mathrm{sh}}$, which means under an ensemble-average power constraint, the VURC ensemble can use a larger support radius than a spherical-shell ensemble with the same average power.

The above average-power construction is included only to clarify the difference between the two power constraints. In the remainder of this paper, we adopt the maximum power constraint for each constituent VURC codebook. This choice imposes the power constraint on every codeword realization and also permits a direct comparison with conventional spherical-shell codebooks under the same maximum power budget.

We next distinguish the resulting power constraints for the MAC and BC. For the Gaussian MAC, the constituent codebooks are independently generated at different transmitters, and each user independently
satisfies
\begin{align}
    \|\mathbf x_j(m_j)\|^2
    \leq
    P_j,
    \qquad
    \forall m_j\in[M_j].
    \label{eq:vurc-mac-max-power}
\end{align}
Therefore, the corresponding VURC radius is $A_j=\sqrt{P_j}$. 

For the Gaussian BC, the $K$ constituent VURC layers are independently generated, where each constituent layer still satisfies the layerwise maximum power constraint as in \eqref{eq:vurc-mac-max-power}. Since the VURC distribution is spherically symmetric, we have $\mathbb E[\mathbf X_j]=\mathbf 0$. Moreover, the constituent codebooks
are generated independently so that
\begin{align}
    \mathbb E\left[
        \|\mathbf X\|^2
    \right]
    &=
    \mathbb E\left[
        \left\|
            \sum_{j=1}^{K}\mathbf X_j
        \right\|^2
    \right]\\
    &=
    \sum_{j=1}^{K}
    \mathbb E\left[
        \|\mathbf X_j\|^2
    \right]
    +
    2\sum_{i<j}
    \mathbb E\left[
        \mathbf X_i^{\mathsf T}\mathbf X_j
    \right]\\
    &=
    \frac{n}{n+2}
    \sum_{j=1}^{K}P_j,
    \label{eq:vurc-bc-total-average-power}
\end{align}
where the cross terms vanish because the independently generated VURC layers are zero mean. Accordingly, under a total BC power budget $P$, we adopt the layerwise power allocation $\sum_{j=1}^{K}P_j\leq P$. Then,
\begin{align}
    \mathbb E\left[
        \|\mathbf X\|^2
    \right]
    &=
    \frac{n}{n+2}
    \sum_{j=1}^{K}P_j\\
    &\leq
    \frac{n}{n+2}P
    <
    P.
    \label{eq:vurc-bc-total-power}
\end{align}
Hence, the proposed BC construction employs layerwise maximum power constraints together with an ensemble-average total power constraint on the superposed codeword.

\subsection{Geometric Advantage over Spherical-Shell Codebooks}
\label{subsec:vurc-geometric-advantage}

We next investigate the geometric advantage of the proposed VURC codebook over the conventional spherical-shell codebook under the maximum power constraint. Since both the MAC and BC are constructed from multiple constituent codebooks, the following results apply identically to each constituent codebook in either channel.

For an arbitrary finite codebook $\mathcal C$, define its minimum Euclidean distance as
\begin{align}
    d_{\min}(\mathcal C)
    \triangleq
    \min_{\substack{
        \mathbf x,\mathbf x'\in\mathcal C,\mathbf x\neq\mathbf x'
    }}
    \|\mathbf x-\mathbf x'\|.
    \label{eq:vurc-dmin}
\end{align}
For a prescribed minimum distance $d$, we further define the maximum packing cardinalities under the spherical-shell and VURC constraints as
\begin{align}
    N_{\mathrm{sh}}(n,A,d)
    &\triangleq
    \max
    \left\{
        |\mathcal C|:
        \mathcal C\subseteq\mathcal S_n(A),
        \,
        d_{\min}(\mathcal C)\geq d
    \right\},\\
    N_{\mathrm{VURC}}(n,A,d)
    &\triangleq
    \max
    \left\{
        |\mathcal C|:
        \mathcal C\subseteq\mathcal B_n(A),
        \,
        d_{\min}(\mathcal C)\geq d
    \right\}.
    \label{eq:vurc-packing-number}
\end{align}

The geometric advantage of VURC is allowing codewords to occupy the radial dimension of the power sphere, i.e., the hypersphere can accommodate multiple mutually separated spherical layers as formalized below.

\begin{lemma}
\label{lem:vurc-multishell-packing}
For any $0<d<A$, the maximum packing cardinality inside $\mathcal B_n(A)$ satisfies
\begin{align}
    N_{\mathrm{VURC}}(n,A,d)
    \geq
    \sum_{\ell=0}^{\lfloor A/d\rfloor}
    N_{\mathrm{sh}}
    \left(
        n,A-\ell d,d
    \right).
    \label{eq:vurc-multishell-bound}
\end{align}
Consequently,
\begin{align}
    N_{\mathrm{VURC}}(n,A,d)
    >
    N_{\mathrm{sh}}(n,A,d).
    \label{eq:vurc-strict-packing-gain}
\end{align}
\end{lemma}

\begin{IEEEproof}
Let $ L\triangleq\left\lfloor\frac{A}{d}\right\rfloor$ and define a sequence of concentric radii $A_\ell\triangleq A-\ell d,\ell=0,\ldots,L$. For every $\ell$, choose a spherical codebook $\mathcal C_\ell\subseteq\mathcal S_n(A_\ell)$ satisfying $|\mathcal C_\ell|=N_{\mathrm{sh}}(n,A_\ell,d)$ and $d_{\min}(\mathcal C_\ell)\geq d$. Consider the union $\mathcal C_{\mathrm{multi}}\triangleq\bigcup_{\ell=0}^{L}\mathcal C_\ell$, we have $\mathcal C_{\mathrm{multi}}\subseteq\mathcal B_n(A)$.

For two codewords belonging to the same layer, their distance is no smaller than $d$ by construction. We therefore only need to consider two codewords $\mathbf x\in\mathcal C_\ell$ and $\mathbf y\in\mathcal C_k$ with $\ell\neq k$. By the reverse triangle inequality, we have
\begin{align}
d_{\min}(\mathcal C_{\mathrm{multi}})&\geq
    \|\mathbf x-\mathbf y\|\\
    &\geq
    \left|
        \|\mathbf x\|-\|\mathbf y\|
    \right|\\
    &=
    |A_\ell-A_k|\\
    &=
    |\ell-k|d
    \geq d,
    \label{eq:vurc-cross-layer-distance}
\end{align}
together with
\begin{align}
    |\mathcal C_{\mathrm{multi}}|
    =
    \sum_{\ell=0}^{L}
    N_{\mathrm{sh}}(n,A-\ell d,d),
\end{align}
which proves \eqref{eq:vurc-multishell-bound}.

Since $0<d<A$, we have $L\geq1$. In addition to the outermost spherical codebook corresponding to $\ell=0$, at least one additional radial layer can therefore be accommodated, which establishes \eqref{eq:vurc-strict-packing-gain}. This completes the proof.
\end{IEEEproof}

The above deterministic packing result can be connected to the VURC ensemble through its full support over the power sphere. Consider any finite feasible sphere packing $\mathcal C^\star=\{\mathbf c_1,\ldots,\mathbf c_M\}\subseteq\mathcal B_n(A)$ and an arbitrarily small $\delta>0$. Since the VURC distribution is strictly positive over the interior of $\mathcal B_n(A)$, every set $\mathcal B_n(\mathbf c_m,\delta)\cap\mathcal B_n(A)$ has a strictly positive probability under VURC. Consequently, there exists a VURC realization in which the $m$th codeword lies within distance $\delta$ of $\mathbf c_m$ for all $m$. For such a
realization, the triangle inequality gives
\begin{align}
    d_{\min}(\mathcal C_{\mathrm{VURC}})&=\|\mathbf X_m-\mathbf X_\ell\|\\
    &\geq
    \|\mathbf c_m-\mathbf c_\ell\|-2\delta,\\
    &=
    d_{\min}(\mathcal C^\star)-2\delta.
    \label{eq:vurc-support-packing}
\end{align}
Since $\delta$ can be made arbitrarily small, the support of the VURC ensemble can approach any feasible finite packing inside the power sphere arbitrarily closely. This justify the geometric advantage of the proposed VURC codebook.

\subsection{Enlarged VURC-Advantage Region in Multiuser Channels}
\label{subsec:vurc-multiuser-region}
Lemma~\ref{lem:vurc-multishell-packing} also leads to a useful multiuser interpretation. To strictly characterize the VURC-advantage region, We first recall the kissing-number characterization associated with a spherical-shell codebook \cite{10.5555/39091,KN1}. Let $M_n^\star$ denote the maximum number of codewords that can be placed on an $n$-dimensional spherical shell of radius $A$ while maintaining a minimum Euclidean distance no smaller than $A$. Equivalently, $M_n^\star$ is the kissing number associated with the angular separation $\pi/3$, and, by scale invariance, it depends only on the dimension $n$ and not on $A$. We then define
\begin{align}
    R_n^\star
    \triangleq
    \frac{1}{n}
    \log_2 M_n^\star.
    \label{eq:vurc-threshold-rate}
\end{align}
Accordingly, for a single codebook with $M=2^{nR}$, the condition $M>M_n^\star$, or equivalently $R>R_n^\star$, indicates that the required codebook cardinality exceeds that supported by the corresponding spherical-shell packing with minimum distance $A$.

In a $K$-user MAC or BC, there are $K$ constituent codebooks with $M_j=2^{nR_j}$. Hence, the cardinality of the $j$th constituent codebook exceeds the corresponding kissing number whenever $R_j>R_n^\star$. For a rate vector $\mathbf R=(R_1,\ldots,R_K)$, define the set of constituent codebooks operating in this regime as
\begin{align}
    \mathcal J_{\mathrm V}(\mathbf R)
    \triangleq
    \left\{
        j\in[K]:
        R_j>R_n^\star
    \right\}.
    \label{eq:vurc-advantage-user-set}
\end{align}
Accordingly, at least one constituent codebook admits the strict VURC advantage if and only if $\mathcal J_{\mathrm V}(\mathbf R)\neq\emptyset$, or equivalently, $\max_{j\in[K]}R_j>R_n^\star$. The corresponding rate region that VURC codebook has advantage over spherical shell codebook is therefore
\begin{align}
    \mathcal R_{\mathrm V}^{(1)}
    \triangleq
    \left\{
        \mathbf R\in\mathbb R_+^K:
        \max_{j\in[K]}R_j>R_n^\star
    \right\},
    \label{eq:vurc-multiuser-region}
\end{align}
and its complement is the $K$-dimensional hypercube
\begin{align}
    \left(
        \mathcal R_{\mathrm V}^{(1)}
    \right)^c
    =
    [0,R_n^\star]^K.
    \label{eq:vurc-no-advantage-hypercube}
\end{align}

Equation~\eqref{eq:vurc-no-advantage-hypercube} reveals an important difference between the single-user and multiuser settings. In the single-user case, only one rate dimension is available to cross the threshold $R_n^\star$. In contrast, a $K$-user system has $K$ independent constituent-rate dimensions, and the VURC-advantage regime is entered as soon as any one of them exceeds the threshold. Thus, in the multiuser rate space, the region without a guaranteed VURC advantage is confined to the bounded hypercube in \eqref{eq:vurc-no-advantage-hypercube}, while all rate vectors outside this hypercube possess at least one constituent codebook that benefits from the additional radial degree of freedom.

A particularly useful sufficient condition can be obtained in terms of the sum rate. Define $R_{\Sigma}\triangleq\sum_{j=1}^{K}R_j$. If none of the constituent codebooks operates in the VURC-advantage regime, then $ R_j\leq R_n^\star,\forall j\in[K]$, which necessarily implies $R_{\Sigma}\leq K R_n^\star$. Therefore, whenever the aggregate coding rate exceeds $K R_n^\star$, at least one constituent codebook must operate beyond the spherical-shell packing threshold and consequently admits the strict geometric advantage established in Lemma~\ref{lem:vurc-multishell-packing}.

\section{Main Results}

In this section, we present three groups of results: i) a closed-form achievability bound and a corresponding lower bound on the average error probability under SIC decoding; ii) numerical and closed-form achievability bounds under JNN decoding; and iii) a series of asymptotic bounds under SIC decoding.

\subsection{Two Bounds under SIC}
\label{section ach}
First, we present a closed-form achievability bound under SIC decoding in the following theorem.

\begin{theorem}
\label{ach}
For the BC and MAC defined in \eqref{BCd} and \eqref{MACd}, under SIC decoding, a closed-form achievability bound for the average error probability for user $i$ is given by
\begin{align}
    \mathbb{E}\{\epsilon_i\}&\leq\sum_{j=1}^{i}\mathbb{E}\{\tilde{\epsilon}_{j}\},\label{ach-1}
\end{align}
where $\tilde{\epsilon_j}$ is the error probability incurred in decoding the codeword of user $j$ given that all codewords of the previous users, i.e., user 1 to $j-1$ (if exists), have been successfully decoded. $\mathbb{E}\{\tilde{\epsilon}_{j}\}$ is bounded by
\begin{align}
     \mathbb{E}\{\tilde{\epsilon}_{j}\}&\leq M_j\,\mathcal{A}_n^{\text{SIC}}(\varepsilon_j),\label{ach-2}
\end{align}
where
\begin{align}
    \varepsilon_{j}&\triangleq\begin{cases}\frac{h_jP_j}{\sigma_{j}^2+h_j\sum_{m=j+1}^{K}P_m},&\text{BC}\\
    \frac{h_jP_j}{\sigma^2+\sum_{m=j+1}^{K}h_mP_m},&\text{MAC},\end{cases}\label{ach-sinr}
\end{align}
and
\begin{align}
    \mathcal A_n^{\text{SIC}}(\varepsilon)
    \triangleq&
    \left(
        \frac{n}{2}\varepsilon
    \right)^{-\frac n2}
    \frac{
        \gamma\left(
            n,\frac n2\varepsilon
        \right)
    }{
        \Gamma\left(\frac n2\right)
    }
    +
    \frac{
        \Gamma\left(
            \frac n2,\frac n2\varepsilon
        \right)
    }{
        \Gamma\left(\frac n2\right)
    }.
    \label{ach-A}
\end{align}

\end{theorem}

\noindent\textit{Proof:} The proof of this achievability bound consists of four main steps. First, the overall error event for each user is decomposed into the sum of the errors from each decoding step using the union bound. Second, the error event of each decoding step is modeled, and its corresponding error probability is derived. Third, the interference-plus-noise vector of each decoding step is statistically replaced by a carefully-designed isotropic Gaussian vector. Finally, a globally convex radial upper bound on the conditional error probability is constructed, which allows the Gaussian replacement to preserve the \textit{achievability} property of the bound. Details of the derivation are placed in Appendix~\ref{t1}. $\hfill\square$

Compared to the proposed closed-form result, it is important to note that the results derived under the CLT regime, e.g., \cite{7300429}, \cite{11006123}, and \cite{9535162}, are not closed-form, due to the term
\begin{align}
    Q^{-1}(\epsilon,\Sigma)\triangleq\left\{\mathbf{v}\in\mathbb{R}^{2^K-1}\Big|\mathrm{Pr}\{\mathbf{u}<\mathbf{v}\}\geq1-\epsilon\right\}, \label{achr-1}
\end{align}
where $\mathbf{u}\sim\mathcal{N}(\mathbf{0},\Sigma)$ is a $K$-dimensional Gaussian random variable. The term in \eqref{achr-1} can only be evaluated numerically via iterative search, rather than expressed analytically. In addition, the proposed achievability bound exhibits substantially lower computational complexity. Existing results have a computational complexity of $O(2^K)$, which grows exponentially with the number of users since the dimension of the associated matrix is $(2^K-1)$ as shown in \eqref{achr-1}. The proposed achievability bound has a polynomial complexity of $O(K^2)$, as indicated in \eqref{ach-1}. Thus the proposed closed-form bound is significantly more computationally efficient, particularly when the number of users $K$ grows large.

Although not explicitly defined in the theorem, $\varepsilon_j$ represents the equivalent SINR for user $j$. Interestingly, this equivalent SINR in \eqref{ach-sinr} is \textit{exactly identical} to that in the infinite blocklength case. It is worth emphasizing that this result is \textit{not} artificially constructed, as it naturally emerges from the multivariate convex-order construction used for the Gaussian replacement. This finding implies that the interference distribution deteriorates for decoding as the blocklength increases, and the interference experienced under infinite blocklength represents the worst-case Gaussian scenario as promised by the classical information theory. Notably, this observation contrasts with the common intuition that finite blocklength generally degrades decoding performance. However, the performance advantage brought by the more favorable interference distribution at finite blocklengths is marginal compared with the inherent impact of channel dispersion due to the finite blocklength.

The closed-form result can be written in the form of error exponent with some approximations to gamma functions. For reasonably large $n$, e.g., $n>20$, we have
\begin{align}
\frac{\gamma\left(n,\frac{n}{2}\varepsilon_{j}\right)}{\Gamma\left(\frac{n}{2}\right)}&=\frac{\gamma\left(\frac{n}{2},\frac{n}{2}\varepsilon_{j}\right)}{\Gamma\left(\frac{n}{2}\right)}\frac{\Gamma(n)}{\Gamma\left(\frac{n}{2}\right)}\\
&=\left[ 1+O\!\left(n^{-1/2}e^{-n}\right)\right]\cdot\left\{ \frac{1}{\sqrt{2}}\left(\frac{2n}{e}\right)^{n/2}\left[1+O\!\left(n^{-1}\right)\right]\right\}\\
&= \frac{\sqrt{2}}{2}\exp\left\{-\frac{n}{2}\left[1-\ln 2n\right]\right\}\cdot\left[1+O\left(n^{-1}\right)\right]\label{ap-1},\\
\frac{\Gamma\left(\frac{n}{2},\frac{n}{2}\varepsilon_{j}\right)}{\Gamma\left(\frac{n}{2}\right)}&=\frac{
        \exp\!\left\{
            -\frac n2
            \left(
                \varepsilon_j-1-\ln\varepsilon_j
            \right)
        \right\}}
        {(\varepsilon_j-1)\sqrt{\pi n}}
    \left[
        1+O\!\left(n^{-1}\right)
    \right],\label{ap-2}
\end{align}
where \eqref{ap-1} employs $\Gamma(z)\approx\sqrt{2\pi}z^{z-\frac{1}{2}}\mathrm{e}^{-z}$ and \eqref{ap-2} employs Stirling expansion. Therefore, we have
\begin{align}
    \mathbb{E}\left\{\tilde{\epsilon}_{j}\right\}
    &\triangleq\exp\left\{-n\left[E_0(\varepsilon_j)-R_j+O\left(n^{-1}\right)\right]\right\},\label{ap-12}
\end{align}
where $E_0(\varepsilon_j)\triangleq\frac12\ln\frac{e\varepsilon_j}{4}$ and $R_j=\frac{1}{n}\log M_j$ is the channel coding rate for user $j$. From \eqref{ap-2}, the average error probability for all users decreases exponentially with the blocklength. This result is consistent with that of the single-user case, as the form of \eqref{ap-12} matches that of Gallager's error exponent \cite{SP67}, indicating that the proposed finite-blocklength bound for the MAC and BC preserves the same large-deviation mechanism underlying the classical error exponent analysis.

Theoretically, a tighter achievability bound can be obtained by directly employing the single-user numerical achievability bound given in \cite[Th.~1]{TIT} 
\begin{equation}
    \tilde{\epsilon}_j\leq1-\int_{0}^{\infty} \int_{0}^{\sqrt{h_jP_j}+w}\left(1-\tilde{\epsilon}_{j1|\mathbf{n}_j}\right)^{M-1}f_{Y|W}(y|w)f_W(w)\,{\rm d} y\, {\rm d} w,\label{2-7}
\end{equation}
where $\tilde{\epsilon}_{j1|\mathbf{n}_j}$ is given in \eqref{2-5} and \eqref{2-6}, $f_{Y|W}(y|w)$ is given in \eqref{2-4}, and
\begin{align}
    f_W(w)&=\frac{1}{\sigma_{j,N1}}\frac{1}{2^{\frac{n}{2}-1}\Gamma(\frac{n}{2})}\left(\frac{w}{\sigma_{j,N1}}\right)^{n-1}\mathrm{e}^{-\frac{w^2}{2\sigma_{j,N1}^2}},\\
    \sigma_{j,N1}^2&=\begin{cases}\frac{1}{n}\left(\sigma_{j}^2+h_j\sum_{m=j+1}^{K}P_m\right),&\text{for the BC}\\
    \frac{1}{n}\left(\sigma^2+\sum_{m=j+1}^{K}h_mP_m\right),&\text{for the MAC}.\end{cases}\label{s1}
\end{align}
The remaining task is to prove that $\tilde{\epsilon}_{j1|\mathbf{n}_j}$ in \eqref{2-7} is convex, thereby satisfying the necessary condition for the SOSD required to establish achievability. Numerical evaluations confirm that $\tilde{\epsilon}_{j1|\mathbf{n}_j}$ is indeed convex, however, establishing this result analytically remains challenging.

Based on the intermediate results of the achievable result in Theorem~1, we further derive a lower bound on the average error probability $\mathbb{E}\{\epsilon_i\}$ in the following.

\begin{corollary}
    \label{conv}
For the BC and MAC defined in \eqref{BCd} and \eqref{MACd}, under SIC decoding, a lower bound on the average error probability of user $i$ is given by
\begin{align}
    \mathbb{E}\{\epsilon_i\}&\geq\mathbb{E}\{\tilde{\epsilon}_{i}\},\label{conv-0}
\end{align}
and
\begin{align}
    \mathbb{E}\{\tilde{\epsilon}_{i}\}\geq\int_0^{\infty}\hat{\epsilon}_{j1|W}f_W(w)\,\dd w,\label{conv-3}
\end{align}
where
\begin{align}\hat{\epsilon}_{j1|W}&\triangleq\hat{\epsilon}_{j1|W}^{11}+\hat{\epsilon}_{j1|W}^{12}+\hat{\epsilon}_{j1|W}^{13}+\hat{\epsilon}_{j1|W}^{21}+\hat{\epsilon}_{j1|W}^{22}+\hat{\epsilon}_{j1|W}^{23},\label{conv-1}\\
    \hat{\epsilon}_{j1|W}^{11}&=\left(\frac{w}{\sqrt{h_jP_j}}-\frac{w^2}{h_jP_j}\right)^n,\label{conv-101}\\
    \hat{\epsilon}_{j1|W}^{12}&=\frac{1}{4}\left(\frac{w}{\sqrt{h_jP_j}}\right)^n\left[(h_jP_j-w^2)^{\frac{n}{2}}-\left(\sqrt{h_jP_j}-w\right)^n\right],\\
    \hat{\epsilon}_{j1|W}^{13}&=\frac{n}{n^2-1}B^{-1}\left(\frac{n+1}{2},\frac{1}{2}\right)B^{-1}\left(\frac{n-1}{2},\frac{1}{2}\right)\frac{(h_jP_j)^{n-\frac{1}{2}}}{4^n}\frac{1}{w^{n-1}}\int_{\sqrt{h_jP_j-w^2}}^{\sqrt{h_jP_j}+w}\frac{P(y)}{y^{n+1}}\,\dd y,\\
    \hat{\epsilon}_{j1|W}^{21}&=\left(\frac{w}{\sqrt{h_jP_j}}-1\right)^n,\\
    \hat{\epsilon}_{j1|W}^{22}&=\frac{1}{4}\left[\left(\frac{w^2}{h_jP_j}-1\right)^{\frac{n}{2}}-\left(\frac{w}{\sqrt{h_jP_j}}-1\right)^n\right],\\
    \hat{\epsilon}_{j1|W}^{23}&=\frac{n}{n^2-1}B^{-1}\left(\frac{n+1}{2},\frac{1}{2}\right)B^{-1}\left(\frac{n-1}{2},\frac{1}{2}\right)\frac{(h_jP_j)^{n-\frac{1}{2}}}{4^n}\frac{1}{w^{n-1}}\int_{\sqrt{w^2-h_jP_j}}^{\sqrt{h_jP_j}+w}\frac{P(y)}{y^{n+1}}\,\dd y,\label{conv-106}\\
    P(y)&=\left(w+y+\sqrt{h_jP_j}\right)\left(-w+y+\sqrt{h_jP_j}\right)\left(w-y+\sqrt{h_jP_j}\right)\left(w+y-\sqrt{h_jP_j}\right),
\end{align}
and
\begin{align}
    f_W(w)=\frac{1}{2^{\frac{n}{2}-1}\sqrt{n}\Gamma(\frac{n}{2})\sigma_{j,N2}}\left(\frac{w}{\sqrt{n}\sigma_{j,N2}}\right)^{n-1}\mathrm{e}^{-\frac{w^2}{2\sigma_{j,N2}^2}}.\label{conv-4}
\end{align}
Here, $\sigma_{j,N2}$ satisfies
\begin{align}
    \int_{\mathcal{D}}\hat{\epsilon}_{j1|\mathbf{n}_j}^{(2)}(\mathbf{n})S_{\hat{\mathbf{n}}_j}(\mathbf{n})\,\dd \mathbf{n}\leq\int_{\hat{\mathcal{D}}}\hat{\epsilon}_{j1|\mathbf{n}_j}^{(2)}(\mathbf{n})S_{\hat{\mathbf{n}}_j^*}(\mathbf{n})\,\dd \mathbf{n},\label{conv-2}
\end{align}
where $\mathcal{D}\triangleq\{\mathbf{n}|S_{\hat{\mathbf{n}}_j}(\mathbf{n})>0\}$, $S_{\hat{\mathbf{n}}_j}(\mathbf{n})\triangleq\int_{\hat{\mathbf{D}}_1}^{\mathbf{n}}\int_{\hat{\mathbf{D}}_1}^{\mathbf{n'}}f_{\hat{\mathbf{n}}_j}(\mathbf{n''})\,\dd\mathbf{n''}\,\dd\mathbf{n'}$, $f_{\hat{\mathbf{n}}_j}(\mathbf{n})$ is the PDF of $\hat{\mathbf{n}}_j\triangleq[\hat{N}_{j1},\hat{N}_{j2},\cdots,\hat{N}_{jn}]$, $\hat{N}_{jl}\sim\mathcal{N}\left(0,\sigma_{j,N2}^2\right)$, and $\hat{\mathbf{D}}_1=\{\mathbf{n}|F_{\hat{\mathbf{n}}_j}(\mathbf{n})=0\}$. Similarly, $\hat{\mathcal{D}}\triangleq\{\mathbf{n}|S_{\mathbf{n}_j}(\mathbf{n})>0\}$, $S_{\hat{\mathbf{n}}_j^*}(\mathbf{n})\triangleq\int_{\hat{\mathbf{D}}_2}^{\mathbf{n}}\int_{\hat{\mathbf{D}}_2}^{\mathbf{n'}}f_{\hat{\mathbf{n}}_j^*}(\mathbf{n}'')\,\dd\mathbf{n''}\,\dd\mathbf{n'}$, $f_{\hat{\mathbf{n}}_j^*}(\mathbf{n})$ is the PDF of $\hat{\mathbf{n}}_j^*\triangleq[\hat{N}_{j1}^*,\hat{N}_{j2}^*,\cdots,\hat{N}_{jn}^*]$, $\hat{N}_{jl}^*\sim\mathcal{N}\left[0,\left(\sigma_{j,N2}^*\right)^2\right]$, and $\hat{\mathbf{D}}_2=\{\mathbf{n}|F_{\hat{\mathbf{n}}_j^*}(\mathbf{n})=0\}$. $\sigma_{j,N2}^*$ satisfies
\begin{align}
    \left(\sigma_{j,N2}^*\right)^2\leq\left\{\begin{array}{cl}
    \frac{1}{n}\sigma_j^2+\frac{1}{n+1}\sum_{m=j+1}^KP_m,&\text{for the BC}\\[6pt]
     \frac{1}{n}\sigma^2+\frac{1}{n+1}\sum_{m=j+1}^KP_m,& \text{for the MAC}.
     \end{array}\right.\label{s2}
\end{align}
\end{corollary}
\noindent\textit{Proof:} The proof of this lower bound consists of two main steps. First, the original interference-plus-noise distribution is replaced by a Gaussian distribution, whose variance is determined by a converse stochastic-dominance argument. This variance is the minimal value required to preserve the validity of the converse. Then, a radial lower bound on the error probability $\tilde{\epsilon}_{i}$ is derived. The detailed derivations are provided in Appendix~\ref{t2}. $\hfill\square$

The value of $\sigma_{j,N2}$ is obtained numerically with computational complexity of $O(N \log N)$, ensuring high computational efficiency. The value difference between the two sides of \eqref{conv-2} increases monotonically with $\sigma_{j,N2}$. Therefore, the maximum value of $\sigma_{j,N2}$ can be efficiently determined using a bisection search method.

A tighter bound, compared to that presented in Corollary~\ref{conv}, can be obtained by substituting \eqref{conv-3} with the existing result in \eqref{2-7} and $f_W(w)$ in \eqref{conv-4}. However, since the convexity of \eqref{2-7} is difficult to establish analytically, we instead use \eqref{conv-101} which is rigorously proven to be convex.

\subsection{Two Bounds under JNN}

We first present a numerical achievability bound for the two-user case under JNN decoding.

\begin{theorem}
\label{thm:jnn-two-user}
Consider a two-user Gaussian MAC in \eqref{MACd} with the VURC codebooks and the JNN decoder in \eqref{eq:MAC-JNN-decoder}. Define
\begin{align}
    p_1(\mathbf u_1,\mathbf z)
    &\triangleq
    \Pr_{\bar{\mathbf U}_1}
    \left\{
        \left\|
            \mathbf z+\mathbf u_1-\bar{\mathbf U}_1
        \right\|
        \leq
        \|\mathbf z\|
        \,\middle|\,
        \mathbf u_1,\mathbf z
    \right\},
    \\
    p_2(\mathbf u_2,\mathbf z)
    &\triangleq
    \Pr_{\bar{\mathbf U}_2}
    \left\{
        \left\|
            \mathbf z+\mathbf u_2-\bar{\mathbf U}_2
        \right\|
        \leq
        \|\mathbf z\|
        \,\middle|\,
        \mathbf u_2,\mathbf z
    \right\},
    \\
    p_{12}(\mathbf u_1,\mathbf u_2,\mathbf z)
    &\triangleq
    \Pr_{\bar{\mathbf U}_1,\bar{\mathbf U}_2}
    \left\{
        \left\|
            \mathbf z+\mathbf u_1+\mathbf u_2
            -\bar{\mathbf U}_1-\bar{\mathbf U}_2
        \right\|
        \leq
        \|\mathbf z\|
        \,\middle|\,
        \mathbf u_1,\mathbf u_2,\mathbf z
    \right\},
    \label{eq:jnn-p-condition}
\end{align}
where $\bar{\mathbf U}_i\sim\mathrm{Unif}[\mathcal B_n(a_i)]$ is an independently generated competing codeword from the same VURC codebook. Then the average joint decoding error probability satisfies
\begin{align}
    \mathbb E\{\epsilon_{\mathrm J}\}
    \leq
    \mathbb E
    \bigg[
        \min
        \bigg\{
            1,\,
            &(M_1-1)p_1
            +(M_2-1)p_2
            +
            (M_1-1)(M_2-1)p_{12}
        \bigg\}
    \bigg].
    \label{eq:jnn-rcu}
\end{align}
Furthermore, $p_1$, $p_2$, and $p_{12}$ can be explicitly written and obtain
\begin{align}
    \mathbb E\{\epsilon_{\mathrm J}\}
    \leq
    \min
    \bigg\{
        1,\,
        &(M_1-1)\bar p_1
        +(M_2-1)\bar p_2
        +
        (M_1-1)(M_2-1)\bar p_{12}
    \bigg\},
    \label{eq:jnn-explicit-bound}
\end{align}
where
\begin{align}
    \bar p_i
    &=
    \int_0^{2\sqrt{h_iP_i}}
    Q\left(
        \frac{\sqrt n\,d}{2\sigma}
    \right)
    f_{D_i}(d)\,\mathrm{d}d,
    \qquad i=1,2,
    \label{eq:jnn-pi}\\
    \bar p_{12}
    &=
    \int_0^{2\sqrt{h_1P_1}}
    \int_0^{2\sqrt{h_2P_2}}
    \int_{-1}^{1}
    Q
    \left(
        \frac{\sqrt n}{2\sigma}
        \sqrt{
            d_1^2+d_2^2+2d_1d_2t
        }
    \right)
    \nonumber\\
    &\qquad\times
    f_{D_1}(d_1)
    f_{D_2}(d_2)
    f_T(t)
    \,\mathrm{d}t\,\mathrm{d}d_2\,\mathrm{d}d_1,
    \label{eq:jnn-p12}
\end{align}
and
\begin{align}
    f_{D_i}(d)
    &=
    \frac{
        nd^{n-1}
    }{
        (h_iP_i)^{n/2}
    }
    I_{1-\frac{d^2}{4h_iP_i}}
    \left(
        \frac{n+1}{2},
        \frac{1}{2}
    \right),
    \qquad
    0\leq d\leq2\sqrt{h_iP_i},
    \label{eq:jnn-distance-pdf}\\
    f_T(t)
    &=
    \frac{\Gamma(n/2)}
    {\sqrt{\pi}\Gamma((n-1)/2)}
    (1-t^2)^{\frac{n-3}{2}},
    \qquad
    -1<t<1.
    \label{eq:jnn-angle-pdf}
\end{align}

\end{theorem}

\begin{IEEEproof}
Under JNN decoding, the competing message pairs are partitioned into three classes corresponding to errors of user~1 only, user~2 only, or both users. Applying the union bound gives \eqref{eq:jnn-rcu}. Conditioned on the Euclidean displacement $\boldsymbol{\delta}$ between the transmitted and competing superposed codewords, the pairwise error probability is $Q\!\left(\sqrt{n}\|\boldsymbol{\delta}\|/(2\sigma)\right)$. Averaging over the corresponding VURC-induced displacement distributions yields $\bar p_1$, $\bar p_2$, and $\bar p_{12}$. The complete derivation is provided in Appendix~\ref{proof of jnn-two-user}.
\end{IEEEproof}

Based on Theorem~\ref{thm:jnn-two-user}, the numerical integrations in \eqref{eq:jnn-pi} and \eqref{eq:jnn-p12} can be further relaxed to obtain a closed-form achievable bound, which is given in the following corollary.

\begin{corollary}
\label{cor:jnn-closed-form}
For the two-user Gaussian MAC with the VURC codebooks and JNN decoder, the average joint decoding error probability is upper bounded by
\begin{align}
    \mathbb E\{\epsilon_{\mathrm J}\}
    \leq
    \min\Big\{1,\,
    &(M_1-1)\mathcal A_n(\gamma_1)
    +(M_2-1)\mathcal A_n(\gamma_2)
    +(M_1-1)(M_2-1)\mathcal A_n(\gamma_{12})
    \Big\},
    \label{eq:jnn-closed-form}
\end{align}
where
\begin{align}
    \mathcal A_n(\gamma)
    \triangleq&
    \left(\frac{n\gamma}{2}\right)^{-\frac n2}
    \frac{
        \gamma\left(n,\frac{n\gamma}{2}\right)
    }{
        \Gamma\left(\frac n2\right)
    }
    +
    \frac{
        \Gamma\left(\frac n2,\frac{n\gamma}{2}\right)
    }{
        \Gamma\left(\frac n2\right)
    },
    \label{eq:jnn-An}
\end{align}
and
\begin{align}
    \gamma_i
    &\triangleq
    \frac{h_iP_i}{\sigma^2},
    \quad i=1,2,\\
    \gamma_{12}
    &\triangleq
    \frac{\max\{h_1P_1,h_2P_2\}}{\sigma^2}
    =
    \max\{\gamma_1,\gamma_2\}.
    \label{eq:jnn-gamma}
\end{align}
\end{corollary}

\begin{IEEEproof}
Conditioned on the Gaussian-noise magnitude $W$, the single-user pairwise error probability is bounded by $p_i\leq\min\{1,(W/a_i)^n\}$, which yields $\bar p_i\leq\mathcal A_n(\gamma_i)$ for $i=1,2$. For the joint error event, upper bounding the convolution density by $1/V_n(a_{12})$, with $a_{12}=\max\{a_1,a_2\}$, similarly gives $\bar p_{12}\leq\mathcal A_n(\gamma_{12})$, where $\gamma_{12}=\max\{\gamma_1,\gamma_2\}$. Substituting these bounds into \eqref{eq:jnn-explicit-bound} yields \eqref{eq:jnn-closed-form}. The complete derivation is provided in Appendix~\ref{proof of jnn-closed-form}.
\end{IEEEproof}

The result in \eqref{eq:jnn-explicit-bound} is an ensemble-average bound over the independently generated VURC codebooks. Therefore, if its right-hand side is no larger than a prescribed $\epsilon$, there exists at least one deterministic realization $(\mathcal C_1,\mathcal C_2)$ whose joint decoding error probability is no larger than $\epsilon$. Moreover, every codeword in this realization satisfies $\|\mathbf x_i(w_i)\|^2\leq P_i$ by construction. Hence the corresponding rate pair $R_i=n^{-1}\log M_i$, $i=1,2$, is achievable.

The VURC-based JNN analysis can, in principle, also be extended to the Gaussian BC following a methodology analogous to that used in deriving the bounds under SIC. For each receiver, once the set of constituent layers to be jointly decoded is specified, the corresponding competing message tuples can again be partitioned according to the subsets of incorrectly decoded layers, and the associated pairwise error events can be characterized analogously.  Compared with the MAC case, however, the BC formulation becomes considerably more cumbersome because different receivers generally decode different subsets of the superposed layers. Since this extension does not introduce a fundamentally different geometric principle from that already illustrated by the two-user MAC, we do not present the explicit VURC+JNN BC result in this paper. 

The principal performance advantage of JNN over SIC is that joint decoding exploits the complete multiuser codeword structure instead of successively converting the remaining users into effective noise. Moreover, JNN avoids the error propagation inherent in successive decoding and jointly accounts for competitors in which both users are decoded incorrectly. The numerical difference between the proposed JNN and SIC bounds is quantified in Section~\ref{sec:numerical_results}.

The improved decoding capability of JNN comes at the cost of substantially higher decoding and analytical complexity. A $K$-user JNN analysis involves error events associated with all $2^K-1$ nonempty subsets of users, and a direct joint search involves $\prod_{i=1}^{K}M_i$ message tuples. In contrast, the SIC-based framework decomposes the multiuser problem into successive effective single-user problems and therefore admits a much more scalable closed-form characterization. Hence, JNN is mainly considered here for two purposes: i) to characterize the performance gain achievable through joint decoding; ii) to enable a controlled comparison with existing results based on spherical-shell codebooks and JNN decoding, thereby isolating the advantage of the proposed VURC codebook.

Theorem~\ref{thm:jnn-two-user} provides a numerically evaluable non-asymptotic bound that retains the exact geometric distributions of the VURC codeword differences. This result preserves more of the finite-dimensional VURC geometry, at the cost of a three-dimensional numerical integration. In contrast, Corollary~\ref{cor:jnn-closed-form} replaces the integrals by explicit upper bounds expressed only through $\mathcal A_n(\cdot)$. This substantially reduces the computational cost and provides a more interpretable expression in terms of the individual SNRs, at the price of an tightness loss. The resulting gap is examined numerically in Section~\ref{sec:numerical_results}.

An interesting observation is that the closed-form JNN bound in Corollary~\ref{cor:jnn-closed-form} has a structure similar to that of the VURC+SIC bound in Theorem.~\ref{ach}. This similarity provides a useful geometric interpretation of the VURC construction across different decoding architectures. The key reason is that, under both SIC and JNN decoding, the relevant pairwise error events can ultimately be interpreted through the intersection between a VURC-supported signal region and a noise-induced detection region. Therefore, $\mathcal A_n(\cdot)$ can be viewed as a common geometric kernel associated with the VURC ensemble, rather than as a consequence of a particular decoding rule.

\subsection{Asymptotic Bounds under SIC}
\label{section asym}
Based on the previous derivations, the following theorem presents new asymptotic bounds for the Gaussian MAC and BC under SIC decoding.

\begin{theorem}
\label{asym}
    For BC and MAC defined in \eqref{BCd} and \eqref{MACd},  under SIC decoding, an asymptotic bound for the average error probability for user $i$ is given by
\begin{align}
    \mathbb{E}\{\epsilon_i\}&=\sum_{j=1}^{i}\left[\mathbb{E}\{\tilde{\epsilon}_{j}\}+O\left(\mathbb{E}\{\tilde{\epsilon}_{j}^2\}\right)\right],\label{asym-1}
\end{align}
where $\mathbb{E}\{\tilde{\epsilon}_{j}\}$ can be replaced with existing single-user asymptotic bounds, e.g., \eqref{2nd}, \eqref{3rd}, and \eqref{4th}, under an equivalent SNR given by 
\begin{align}
    \varepsilon_{j}&=\begin{cases}\frac{h_jP_j}{\sigma_{j}^2+h_j\sum_{m=j+1}^{K}P_m},&\text{for the BC}\\
    \frac{h_jP_j}{\sigma^2+\sum_{m=j+1}^{K}h_mP_m},&\text{for the MAC}.\end{cases}
\end{align}
\end{theorem}

\noindent\textit{Proof:} Based on the proof of Theorem~\ref{ach}, it suffices to show that the equality in \eqref{asym-1} holds and all inequalities in the proof of Theorem~\ref{ach} becomes equalities as $n \to \infty$. The equality in \eqref{asym-1} is established by verifying that the chain of equalities from \eqref{achp-1001} to \eqref{achp-ab2} all hold as $n \to \infty$. Furthermore, by virtue of the CLT, the true noise distribution asymptotically converges to the constructed noise distribution. Consequently, all inequalities derived under the constructed noise model become equalities as $n \to \infty$.
$\hfill\square$

With regard to single-user asymptotic bounds mentioned in Theorem~\ref{asym}, we present examples of various orders, where $R=\log M$ is the channel coding rate of the single-user system, $M$ is the size of the codebook, $C$ is the capacity of the channel, $\epsilon$ is the average decoding error probability, and $\gamma$ is the SNR.

\textit{(2nd-order, Polyanskiy, Poor, and Verd\'u, 2011 \cite{PPV})}
\begin{align}
    R(n,\epsilon)=C(\gamma)-\sqrt{\frac{V(\gamma)}{n}}Q^{-1}(\epsilon)+O(n^{-1}\log n),\label{2nd}
\end{align}
where
\begin{align}
    V(\gamma)&=\frac{\gamma(\gamma+2)}{2(\gamma+1)^2}\log^2\mathrm{e},\\
    Q(x)&=\frac{1}{\sqrt{2\pi}}\int_{x}^{\infty}\exp{{-t^2/2}}\,\dd t.
\end{align}

\textit{(3rd-order, Tan and Tomamichel, 2015 \cite{Tan})}
\begin{align}
    R(n,\epsilon)=C(\gamma)-\sqrt{\frac{V(\gamma)}{n}}Q^{-1}(\epsilon)+\frac{1}{2n}\log n+o(n^{-1}).\label{3rd}
\end{align}

\textit{(4th-order, Bao, Xu, You, and Poor, 2026 \cite{TIT})}
\begin{align}
    \epsilon=A\sqrt{n}\mathrm{e}^{-Bn}+O\left(\frac{\mathrm{e}^{-n}}{\sqrt{n}}\right),\label{4th}
\end{align}
where $A$ and $B$ are terms of only $\gamma$.

In the very short blocklength regime, e.g., $n\leq 100$, higher-order asymptotic analysis becomes important for further narrowing the approximation gap. The second-order approximation captures the dominant finite-blocklength backoff from capacity by the dominant $O(n^{-1/2})$ penalty, but leaves an unresolved remainder of order $O(\log n/n)$. The third- and fourth-order characterizations successively resolve finer components of this residual approximation error. The third-order analysis explicitly characterizes the $O(\log n/n)$ contribution, whereas the fourth-order analysis further resolves the $O(1/n)$ contribution.

The asymptotic bound in Theorem~\ref{asym} offers a highly extensible framework, where existing single-user asymptotic results can be directly applied to the Gaussian MAC and BC. Specifically, by employing successive decoding and adopting a methodology that separates the overall error event into multiple individual sub-events, the joint decoding process in MAC/BC channels can be decomposed into several effectively independent single-user decodings. This decomposition not only simplifies the analytical characterization of system performance but also facilitates a clearer understanding by separately evaluating how individual user performance contributes to the overall system reliability. Moreover, such a structure allows the asymptotic behavior of each user to be directly mapped to the corresponding multiuser setting, thereby bridging the analytical gap between single-user and multiuser regimes.

The proposed asymptotic result also facilitates extensions to both larger channel dimensions and higher orders. As highlighted in the remarks of Theorem~\ref{ach}, the proposed bounds exhibit polynomial computational complexity, in contrast to the exponential complexity of existing methods. This property makes the proposed bounds particularly suitable for scaling to systems with more users. Furthermore, due to the extensible framework of the proposed asymptotic bound, higher-order multiuser results can be obtained directly by deriving the corresponding higher-order single-user asymptotic results. This approach is significantly simpler than conventional multiuser methodologies, e.g., in \cite{ML,YKEMAC,zhoulin}.

\section{Numerical Evaluation}
\label{sec:numerical_results}

\begin{figure*}[t]
    \centering
    \includegraphics[width=0.98\textwidth]
    {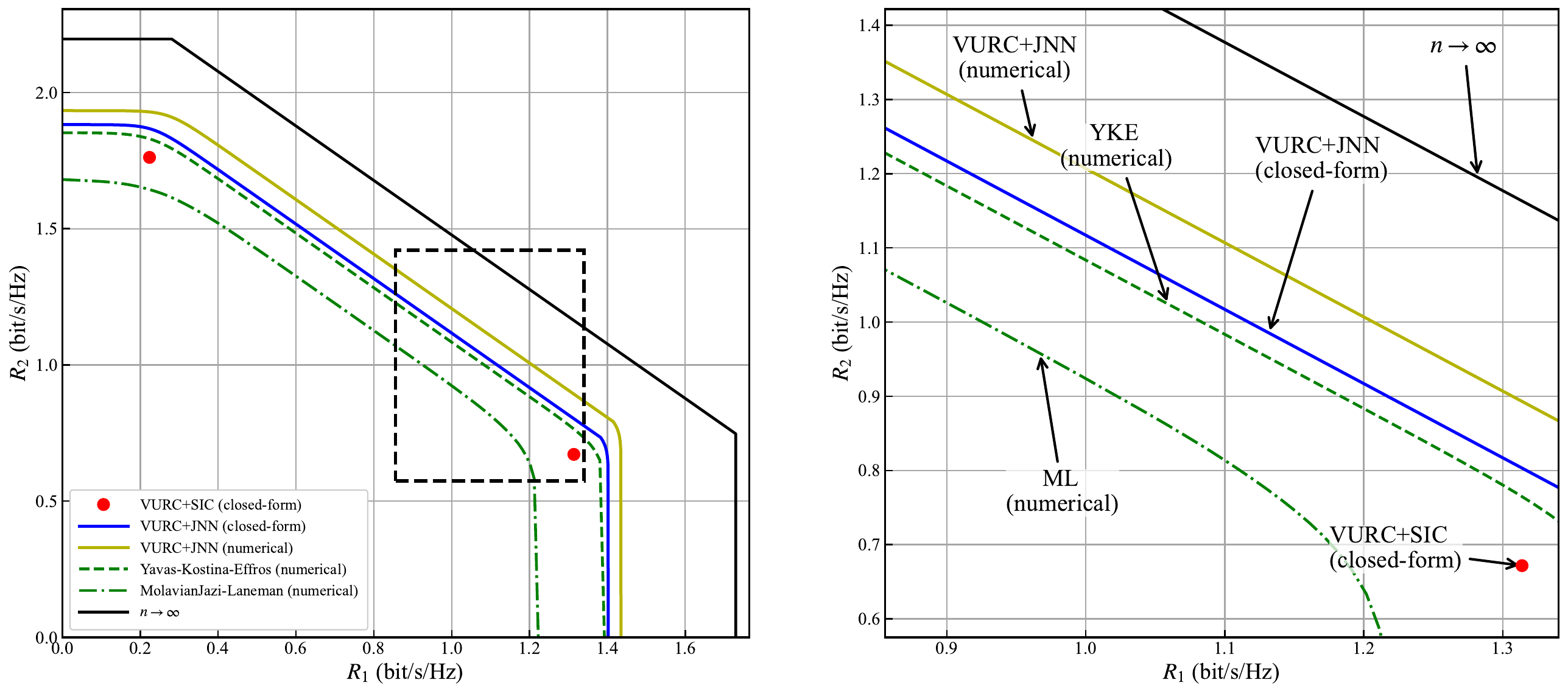}
    \caption{Non-asymptotic achievable rate-region comparison for a
    two-user Gaussian MAC with blocklength $n=50$,
    $\gamma_{1,\mathrm{MAC}}=10$,
    $\gamma_{2,\mathrm{MAC}}=20$,
    $\epsilon_1=2\times10^{-3}$, and
    $\epsilon_2=1\times10^{-3}$. }
    \label{fig:MAC_nonasym}
\end{figure*}

In this section, we numerically evaluate the proposed finite-blocklength framework. We first compare the non-asymptotic achievable results for the Gaussian MAC and BC. We then examine the gap between the SIC achievability bound and the corresponding lower bound on the average error probability. Finally, we investigate the behavior of the proposed higher-order asymptotic characterizations.

We begin with the two-user Gaussian MAC, for which Fig.~\ref{fig:MAC_nonasym} compares the proposed VURC+JNN numerical and closed-form bounds with the VURC+SIC result and the existing non-asymptotic spherical-shell results. As shown in Fig.~\ref{fig:MAC_nonasym}, the VURC+JNN numerical bound provides the largest achievable region among the displayed finite-blocklength bounds in the considered operating regime. This comparison separates two different sources of performance variation: i) the difference between JNN and SIC reflects the decoder architecture, where JNN shows a clear advantage over SIC; ii) the two VURC+JNN bounds provide larger rate region than exisiting spherical shell codebook plus JNN results, demonstating the advantage of the proposed VURC codebook. Since time-sharing can not be applied in finite blocklength regime~\cite{Mross2024}, the VURC+SIC result is represented by the two operating points associated with the two SIC decoding orders.

\begin{figure*}[t]
    \centering
    \includegraphics[width=0.98\textwidth]
    {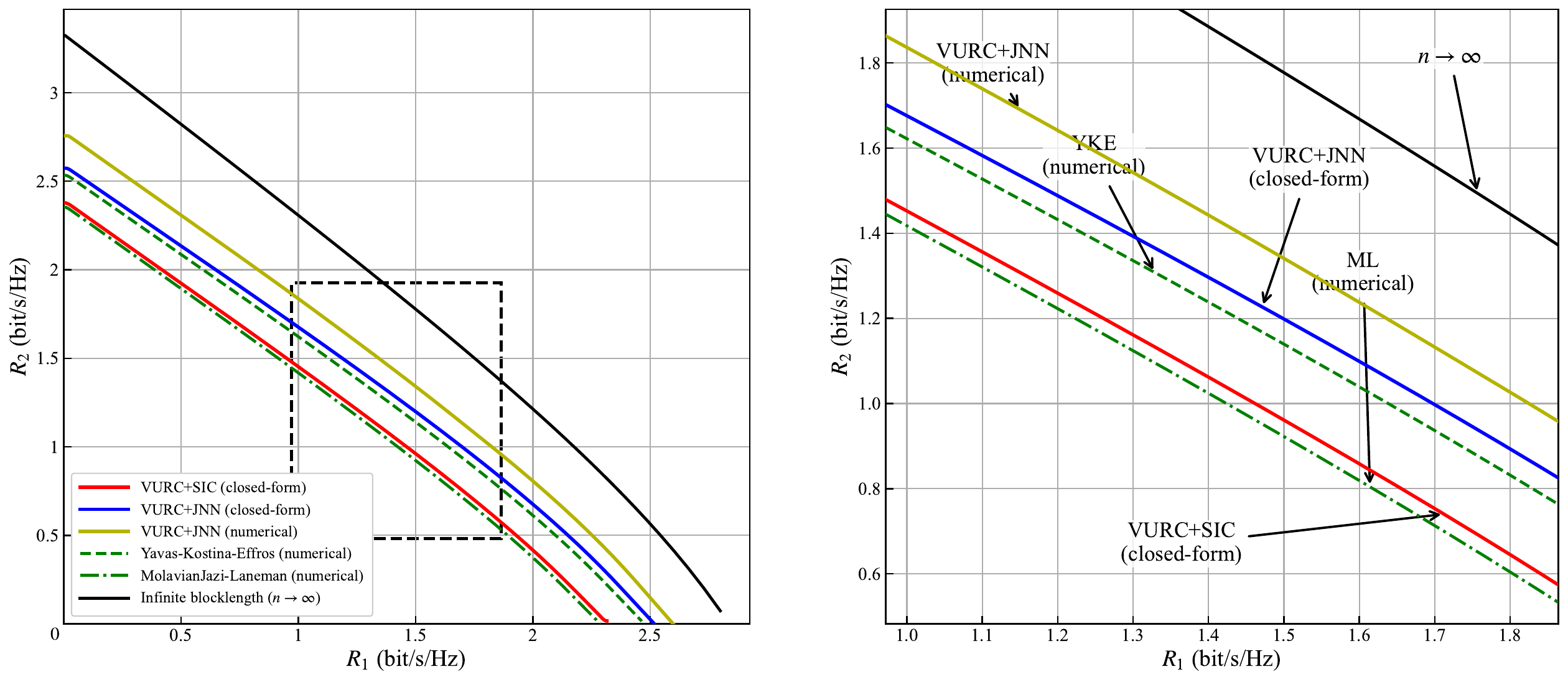}
     \caption{Non-asymptotic achievable rate-region comparison for a
    two-user Gaussian BC with blocklength $n=50$,
    $P_1+P_2=100$, $\sigma_1^2=2$, $\sigma_2^2=1$,
    $\epsilon_1=2\times10^{-3}$, and
    $\epsilon_2=1\times10^{-3}$.}
    \label{fig:BC_nonasym}
\end{figure*}

We next perform an analogous comparison for the Gaussian BC. The rate region is obtained directly from the MAC model by sweeping the feasible constituent-layer power allocations and taking the outer envelope of the resulting rate pairs. To ensure a fair comparison, all schemes adopt the same BC power-control constraint defined in \eqref{eq:BC-layer-power}. As shown in Fig.~\ref{fig:BC_nonasym}, the BC results exhibit similar to that observed for the MAC, where the VURC+JNN numerical bound provides the largest finite-blocklength achievable region, followed by the VURC+JNN closed-form bound.

\begin{figure*}[t]
    \centering
    \includegraphics[width=0.98\textwidth]
    {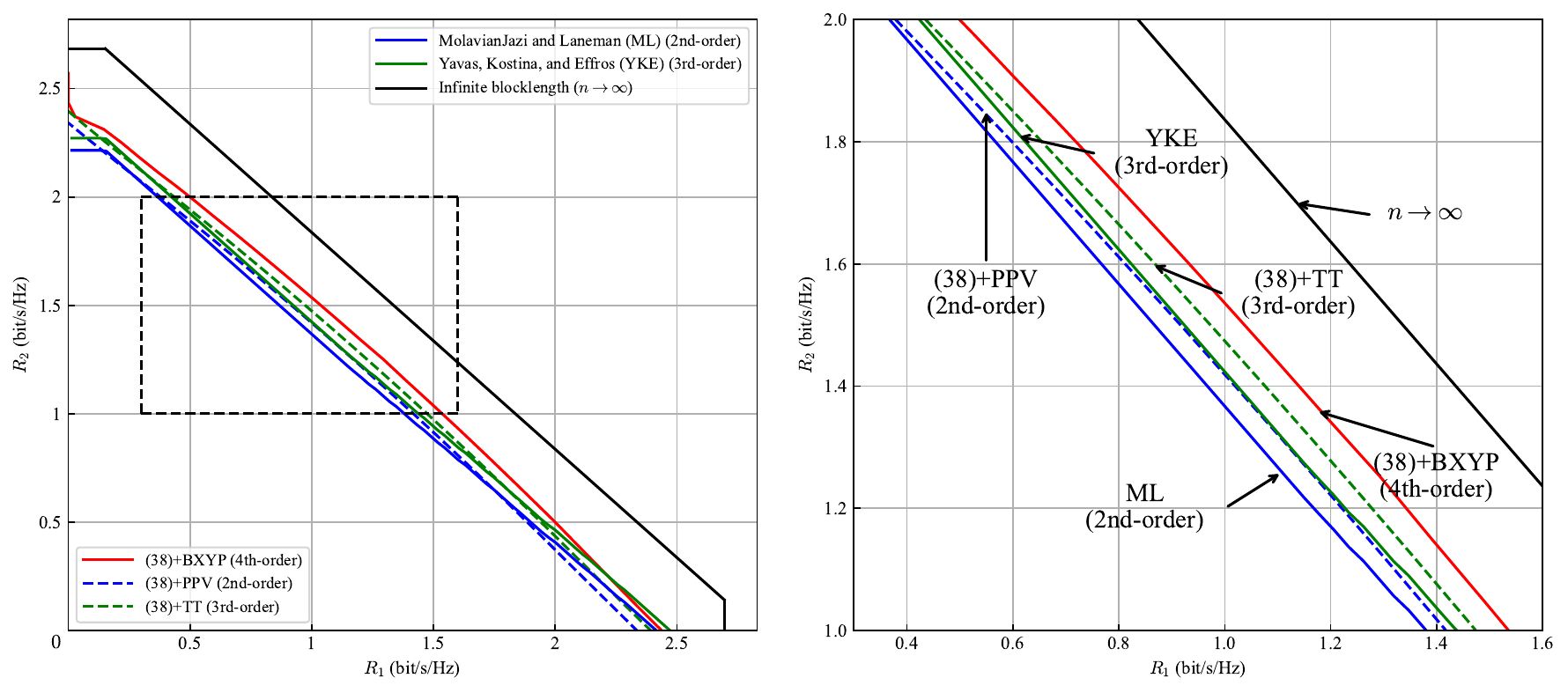}
    \caption{Comparison of asymptotic rate-region characterizations for
    a two-user Gaussian BC with blocklength $n=50$,
    $P_1+P_2=50$, $\sigma_1^2=\sigma_2^2=1$, and
    $\epsilon_1=\epsilon_2=10^{-5}$.}
    \label{fig:BC_asym}
\end{figure*}

Finally, we compare the results obtained by combining Theorem~\ref{asym} with the three single-user asymptotic bounds in \eqref{2nd}, \eqref{3rd}, and \eqref{4th}, along with existing results, as shown in Fig.~\ref{fig:BC_asym}. In general, higher-order results yield tighter achievable rate regions. When comparing these existing results with Theorem~\ref{asym} using the same-order single-user bound, the two sets of results closely align.

\section{Conclusion}
This work established a comprehensive finite-blocklength characterization for the Gaussian MAC and BC by introducing a multiuser VURC codebook design and considering both SIC and JNN decoding. We derived an explicit closed-form non-asymptotic achievability bound under SIC decoding for an arbitrary number of users, together with numerical and closed-form JNN achievability bounds for the two-user Gaussian MAC; a scheme-specific lower bound on the average error probability under SIC decoding; and a series of asymptotic characterizations up to the fourth order. The resulting framework provides analytically interpretable finite-blocklength characterizations while allowing straightforward extensions to more antennas and higher order. The proposed framework can be extended to mismatched scenarios, with potential approaches inspired by the techniques in \cite{Scarlett}.

\begin{appendices}

\section{Proof of Theorem~\ref{ach}}
\label{t1}
In this section, we determine an upper bound of the error probability of each user $\epsilon_i$, given the blocklength $n$, the codebook size for each user $M_i$, and the channel power gain $h_i$. We first derive the error probability of each user, given the magnitude of the interference-plus-noise $\mathbf{n}_j$. Then, we constructed a new interference-plus-noise $\mathbf{\tilde{n}}_j$ and prove that $\mathbb{E}_{\mathbf{n}_j}\{\tilde{\epsilon}_{j|\mathbf{n}_j}\}\leq\mathbb{E}_{\mathbf{\tilde{n}}_j}\{\tilde{\epsilon}_{j|\mathbf{n}_j}\}$. Finally, we prove \eqref{ach-1}. Here, only results for the MAC are presented, while the corresponding results for the BC are derived analogously.

\subsection{Sphere-Packing Model for Calculating $\tilde{\epsilon}_{j|\mathbf{n}_j}$}

According to the definitions, we have
\begin{align}
\mathbf{n}_j=\sum_{m=j+1}^{K}\sqrt{h_m}\mathbf{x}_m+\mathbf{z}.\label{3-4}
\end{align}
Denote $\mathbf{x}_i=[X_{i1},X_{i2},\cdots,X_{in}]$, $\mathbf{z}=[Z_1,Z_2,\cdots,Z_n]$, and $\mathbf{n}_j=[N_{j1},N_{j2},\cdots,N_{jn}]$. Under the onstituent VURC construction, $\mathbf{x}_i$ is uniformly distributed inside a $n$-dimensional hypersphere and is therefore isotropic.

\begin{figure}
\centering
\includegraphics[width=0.8\textwidth]{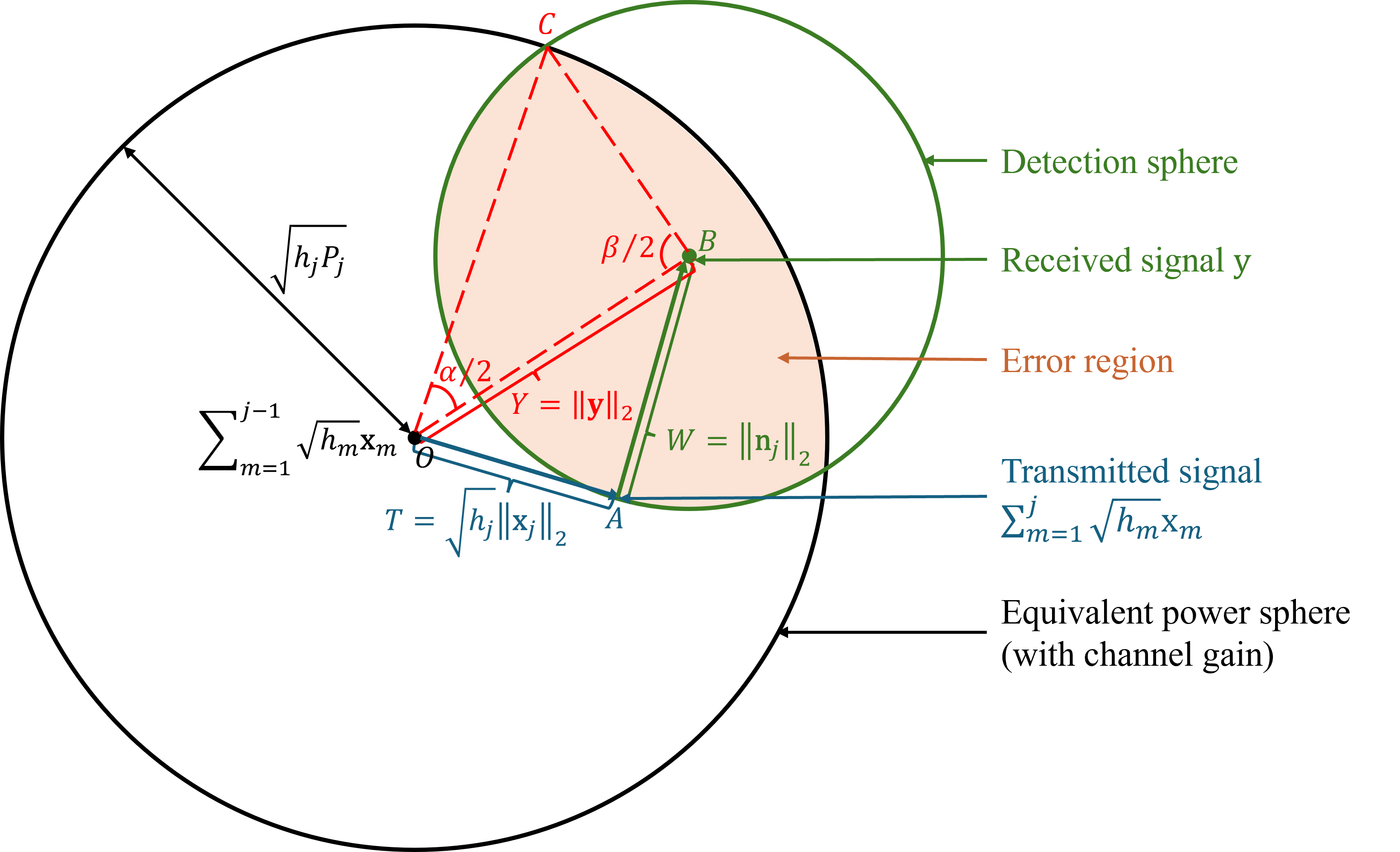}
\caption{Error event analysis for user $j$ using VURC codebook over a Gaussian MAC.}
\label{error region}
\end{figure}

To explicitly write $\tilde{\epsilon_j}$, the corresponding error events need to be analyzed. Fig.~\ref{error region} illustrates the geometrical error event analysis of the proposed codebook over a Gaussian MAC, where node $A$ denotes the equivalent transmitted signal, node $B$ denotes the received signal, the black sphere represents the equivalent power sphere containing all transmitted signals, and the green sphere represents the detection sphere. Detailed descriptions of the equivalent power sphere, the detection sphere, and the error region, are provided in the following.

\textit{Equivalent power sphere}: Consider decoding $\mathbf{x}_j$, given a layerwise maximum power constraint of $P_j$, the equivalent transmitted signal distributed uniformly inside an $n$-dimensional hypersphere centered at $\sum_{m=1}^{j-1}\sqrt{h_m}\mathbf{x}_m$, say node $O$ in Fig.~\ref{error region}, at a radius of $\sqrt{h_jP_j}$. We define this sphere as the \textit{equivalent power sphere}.

\textit{Detection sphere}: According to the definition of the error event of $\tilde{\epsilon_j}$, codewords from $\mathbf{x}_1$ to $\mathbf{x}_{j-1}$ are assumed to have been decoded correctly, which means the remaining noise for decoding $\mathbf{x}_j$ is only $\mathbf{n}_j$. Denote the magnitude of $\mathbf{n}_j$ by a random variable $W$, i.e., $W\triangleq\Vert\mathbf{n}_j\Vert$. With minimum distance decoding, conditioned on $W=w$, if any transmitted signal from the user $j$'s codebook is located closer to the received signal at node $A$ than $w$, an error occurs in the detection. In other words, if any signal other than the truly transmitted one lies inside the sphere with the received signal as the origin and radius $w$, an error occurs in the detection. We define this sphere as the \textit{detection sphere}. 

\textit{Error region}: Since we consider codewords distributed uniformly inside the power sphere, the intersection between the detection sphere and the power sphere (if it exists) contributes to the error probability. We define this intersection as the \textit{error region}. Note that this error region is conditioned on $Y=y$ and $W=w$, where $Y$ is defined as a random variable to characterize the magnitude of the received signal and $Y$ stochastically depends on $W$. The concept of error region is only designed to facilitate the analysis of the decoding error probability. The actual decoding process just determines the closest codeword and so the concept of error region is \textit{not} needed.

Based on this geometric model, we derive the error probability. Denote $\tilde{\epsilon}_{j1|\mathbf{n}_j}$ as the probability of the event that a single codeword, say codeword $\mathbf{x}_j(1)$ without loss of generality, lies in the error region; then we have
\begin{equation}
    \tilde{\epsilon}_{j1|\mathbf{n}_j}=\frac{V_n^{\text{cap}}(w,\beta)+V_n^{\text{cap}}(\sqrt{h_jP_j},\alpha)}{V_n(\sqrt{h_jP_j})},\label{a6-1}
\end{equation}
where $V_n(r)$ denotes the volume of an $n$-dimensional sphere with radius $r$,  $V_n^{\text{cap}}(r,\theta)$ denotes the volume of an $n$-dimensional spherical cap with radius $r$ and polar angle $\theta$, $\alpha=2\angle COB$ and $\beta=2\angle CBO$ are depicted in Fig.~\ref{error region}. By employing
\begin{align}
    V_n^{\text{cap}}(r,\theta)=\left\{\begin{array}{cl}
    V_n(r)\frac{1}{2}I_{\sin^2\theta}(\frac{n+1}{2},\frac{1}{2}),&\theta\leq\frac{\pi}{2}\\
    V_n(r)\left(1-\frac{1}{2}I_{\sin^2\theta}(\frac{n+1}{2},\frac{1}{2})\right),&\theta>\frac{\pi}{2},\\
    \end{array}\right.
\end{align}
we have
 \begin{equation}
\begin{aligned}
     \tilde{\epsilon}_{j1|\mathbf{n}_j}=\left\{\begin{array}{cl}
    \left(\frac{w}{\sqrt{h_jP_j}}\right)^n,&0\leq y\leq \sqrt{h_jP_j}-w\\[6pt]
    \left[1-\frac{1}{2}I_{\sin^2\beta}\left(\frac{n+1}{2},\frac{1}{2}\right)\right] \left(\frac{w}{\sqrt{h_jP_j}}\right)^n+\frac{1}{2}I_{\sin^2\alpha}\left(\frac{n+1}{2},\frac{1}{2}\right),&\sqrt{h_jP_j}-w<y\leq \sqrt{h_jP_j-w^2}\\[6pt]
    \frac{1}{2}I_{\sin^2\beta}\left(\frac{n+1}{2},\frac{1}{2}\right)\left(\frac{w}{\sqrt{h_jP_j}}\right)^n+\frac{1}{2}I_{\sin^2\alpha}\left(\frac{n+1}{2},\frac{1}{2}\right),& \sqrt{h_jP_j-w^2}<y\leq \sqrt{h_jP_j}+w,
     \end{array}\right.\label{2-5}
\end{aligned}
\end{equation}
for $w<\sqrt{h_jP_j}$, and
    \begin{equation}
\begin{aligned}
     \tilde{\epsilon}_{j1|\mathbf{n}_j}=\left\{\begin{array}{cl}
    1,&0\leq y\leq w-\sqrt{h_jP_j}\\[6pt]
    \frac{1}{2}I_{\sin^2\beta}\left(\frac{n+1}{2},\frac{1}{2}\right) \left(\frac{w}{\sqrt{h_jP_j}}\right)^n+\left[1-\frac{1}{2}I_{\sin^2\alpha}\left(\frac{n+1}{2},\frac{1}{2}\right)\right],&w-\sqrt{h_jP_j}<y\leq \sqrt{w^2-h_jP_j}\\[6pt]
     \frac{1}{2}I_{\sin^2\beta}\left(\frac{n+1}{2},\frac{1}{2}\right)\left(\frac{w}{\sqrt{h_jP_j}}\right)^n+\frac{1}{2}I_{\sin^2\alpha}\left(\frac{n+1}{2},\frac{1}{2}\right),& \sqrt{w^2-h_jP_j}<y\leq \sqrt{h_jP_j}+w,
     \end{array}\right.\label{2-6}
\end{aligned}
\end{equation}
for $w\geq\sqrt{h_jP_j}$. Note that $\tilde{\epsilon}_{j1}$ in fact depends not only on $\mathbf{n}_j$, but also on $Y$. However, since an upper bound of $\tilde{\epsilon}_{j1}$ that is independent of $Y$ will be derived, we omit $Y$ from the subscript here for simplicity.

Finally, by applying the union bound, we have
\begin{align}
    \tilde{\epsilon}_{j|\mathbf{n}_j}&\leq1-\left(1-\tilde{\epsilon}_{j1|\mathbf{n}_j}\right)^{M_j-1}\label{achp-1002}\\
    &\leq M_j\tilde{\epsilon}_{j1|\mathbf{n}_j}\label{achp-1003},
\end{align}
where $M_j=|\mathcal{C}_j|$ is the number of codewords in user $j$'s codebook.

\subsection{Construction of $\mathbf{\tilde{n}}_j$ and Vector Stochastic Dominance}
To calculate $\tilde{\epsilon}_j$, we need to average the conditional error probability over the interference-plus-noise vector $\mathbf{n}_j$. According to \eqref{3-4}, its joint distribution is difficult to obtain in closed form because each VURC codeword is uniformly distributed over a hypersphere and hence has dependent coordinates. To avoid any coordinatewise independence assumption, we construct an isotropic Gaussian vector
\begin{align}
    \tilde{\mathbf{n}}_j\sim\mathcal{N}\left(\mathbf{0},\sigma_{j,N1}^2\mathbf{I}_n\right),
\end{align}
and compare the two random vectors in the multivariate convex order.

\begin{theorem}
\label{2nd domin}
For the MAC, let $\mathbf{n}_j$ be defined in \eqref{3-4}. If
\begin{align}
    \sigma_{j,N1}^2\geq\frac{1}{n}\left(\sigma^2+\sum_{m=j+1}^{K}h_mP_m\right),\label{sigmaN1-cx}
\end{align}
then
\begin{align}
    \mathbf{n}_j\preceq_{\rm cx}\tilde{\mathbf{n}}_j.\label{vector-cx-noise}
\end{align}
Consequently, for every jointly convex function $a:\mathbb{R}^n\to\mathbb{R}$ for which the expectations exist,
\begin{align}
    \mathbb{E}_{\mathbf{n}_j}\{a(\mathbf{n}_j)\}
    \leq
    \mathbb{E}_{\tilde{\mathbf{n}}_j}\{a(\tilde{\mathbf{n}}_j)\}.\label{ach-vector-cx}
\end{align}
The corresponding BC result follows analogously after replacing the effective interference and noise powers by those in \eqref{ach-sinr}.
\end{theorem}
\noindent\textit{Proof:} Please refer to Appendix~\ref{2nd domin proof}. $\hfill\square$

The following elementary composition result is used to connect this vector condition to the radial error-probability bounds employed below.

\begin{lemma}
\label{radial convex lemma}
Let $g:[0,\infty)\to\mathbb{R}$ be convex and non-decreasing. Then $a(\mathbf{x})\triangleq g(\Vert\mathbf{x}\Vert)$ is jointly convex in $\mathbf{x}$.
\end{lemma}
\noindent\textit{Proof:}
For arbitrary $\mathbf{x},\mathbf{y}\in\mathbb{R}^n$ and $\lambda\in[0,1]$, the triangle inequality gives
\begin{align}
\Vert\lambda\mathbf{x}+(1-\lambda)\mathbf{y}\Vert
\leq \lambda\Vert\mathbf{x}\Vert+(1-\lambda)\Vert\mathbf{y}\Vert.
\end{align}
Since $g$ is non-decreasing and convex, following Jensen's inequality, we have
\begin{align}
 g\!\left(\Vert\lambda\mathbf{x}+(1-\lambda)\mathbf{y}\Vert\right)
 &\leq g\!\left(\lambda\Vert\mathbf{x}\Vert+(1-\lambda)\Vert\mathbf{y}\Vert\right)\\
 &\leq \lambda g(\Vert\mathbf{x}\Vert)+(1-\lambda)g(\Vert\mathbf{y}\Vert),
\end{align}
which proves the joint convexity. $\hfill\square$

\subsection{Derivation of the Upper Bound of $\mathbb{E}\{\tilde{\epsilon}_{j}\}$}
\label{achp-ss}

By leveraging the geometric relationships in Fig.~\ref{error region}, from \eqref{a6-1}, we have
\begin{align}
    \tilde{\epsilon}_{j1|\mathbf{n}_j}(\mathbf{n}_j)
    &\leq
    \min\left\{
        1,
        \left(
            \frac{W}{\sqrt{h_jP_j}}
        \right)^n
    \right\}
    \nonumber\\
    &=
    \left\{
    \begin{array}{cl}
    \left(
        \dfrac{W}{\sqrt{h_jP_j}}
    \right)^n,
    &0\leq W\leq\sqrt{h_jP_j}\\[6pt]
    1,
    &W>\sqrt{h_jP_j},
    \end{array}
    \right.
    \label{radial-capped}
\end{align}
where $W=\Vert\mathbf{n}_j\Vert$. Applying the Gaussian replacement in Theorem~\ref{2nd domin}, we obtain
\begin{align}
    \mathbb{E}\{\tilde{\epsilon}_{j}\}
    &\leq
    M_j
    \mathbb{E}\left\{
        \min\left[
            1,
            \left(
                \frac{\tilde W}{\sqrt{h_jP_j}}
            \right)^n
        \right]
    \right\}
    \nonumber\\
    &=
    M_j
    \left[
    \frac{1}{(h_jP_j)^{n/2}}
    \int_{0}^{\sqrt{h_jP_j}}
        w^n f_{\tilde W}(w)\,\dd w
    +
    \int_{\sqrt{h_jP_j}}^{\infty}
        f_{\tilde W}(w)\,\dd w
    \right],
    \label{achp-ab1}
\end{align}
where $\tilde W\triangleq\Vert\tilde{\mathbf n}_j\Vert$ follows a scaled $\chi$ distribution with PDF
\begin{align}
    f_{\tilde W}(w)
    =
    \frac{1}{\sigma_{j,N1}}
    \frac{1}
    {2^{\frac n2-1}\Gamma(\frac n2)}
    \left(
        \frac{w}{\sigma_{j,N1}}
    \right)^{n-1}
    \exp\left(
        -\frac{w^2}{2\sigma_{j,N1}^2}
    \right),
    \qquad w\geq0.
    \label{chi-correct}
\end{align}
For the tightest Gaussian replacement allowed by \eqref{sigmaN1-cx}, we choose equality in \eqref{sigmaN1-cx}.
Let
\begin{align}
    x_j
    \triangleq
    \frac{h_jP_j}{2\sigma_{j,N1}^2}
    =
    \frac{n\varepsilon_j}{2}.
\end{align}
Then, by the change of variable $t=w^2/(2\sigma_{j,N1}^2)$, we obtain
\begin{align}
    \frac{1}{(h_jP_j)^{n/2}}
    \int_{0}^{\sqrt{h_jP_j}}
        w^n f_{\tilde W}(w)\,\dd w
    &=
    \left(
        \frac{n}{2}\varepsilon_j
    \right)^{-\frac n2}
    \frac{
        \gamma\left(
            n,\frac n2\varepsilon_j
        \right)
    }{
        \Gamma\left(\frac n2\right)
    },\\
    \int_{\sqrt{h_jP_j}}^{\infty}
        f_{\tilde W}(w)\,\dd w
    &=
    \frac{
        \Gamma\left(
            \frac n2,\frac n2\varepsilon_j
        \right)
    }{
        \Gamma\left(\frac n2\right)
    }.
\end{align}
Therefore,
\begin{align}
    \mathbb{E}\{\tilde{\epsilon}_{j}\}
    &\leq
    M_j\,\mathcal A_n^{\text{SIC}}(\varepsilon_j),
    \label{ach-single-final}
\end{align}
where $\mathcal A_n^{\text{SIC}}$ is given in \eqref{ach-A}.

\subsection{Derivation for $\mathbb{E}\{\epsilon_i\}$}
 We first bound the overall decoding error probability of user $i$ by the error probabilities incurred at the successive SIC stages. According to the SIC decoding process, we have
\begin{align}
    \mathbb{E}\{\epsilon_i\}
    &\leq
    \mathbb{E}\left\{
        1-
        \prod_{j=1}^{i}
        \left(
            1-\tilde{\epsilon}_j
        \right)
    \right\}
    \label{achp-1001}\\
    &\leq
    \mathbb{E}\left\{
        \sum_{j=1}^{i}
        \tilde{\epsilon}_j
    \right\}
    \label{achp-ab2-2}\\
    &=
    \sum_{j=1}^{i}
    \mathbb{E}\left\{
        \tilde{\epsilon}_j
    \right\},
    \label{achp-ab2}
\end{align}
where \eqref{achp-1001} follows because an error in decoding user $i$ can occur only if at least one of the first $i$ SIC decoding stages fails, and \eqref{achp-ab2-2} follows from $1-\prod_{j=1}^{i}(1-x_j)\leq\sum_{j=1}^{i}x_j$ for $0\leq x_j\leq1$.

Finally, by substituting \eqref{ach-single-final} into \eqref{achp-ab2}, we obtain \eqref{ach-1} together with \eqref{ach-2}. This completes the proof for Theorem~\ref{ach}.

\section{Proof of Corollary~\ref{conv}}
\label{t2}

In this section, we determine a lower bound of the error probability $\epsilon_i$ of each user. First, the inequality in \eqref{conv-0} directly establishes since the error event on the right-hand-side is part of the error event on the left-hand-side. The subsequent derivations contain two steps. Due to the similarity between the the MAC and BC, unless otherwise specified, only the results for MAC are presented, while the corresponding results for the BC are omitted for the sake of conciseness.

\subsection{Construction of $\mathbf{\hat{n}}_j$ and Proof of $\mathbb{E}_{\mathbf{n}_j}\{\tilde{\epsilon}_{j|\mathbf{n}_j}\}\geq\mathbb{E}_{\mathbf{\hat{n}}_j}\{\tilde{\epsilon}_{j|\mathbf{n}_j}\}$}

Analogous to the method we employed in the proof of the achievability bound, here we construct a new interference-plus-noise term $\hat{\mathbf{n}}_j\triangleq[\hat{N}_{j1},\hat{N}_{j2},\cdots,\hat{N}_{jn}]$ where $\hat{N}_{j1}$, $\hat{N}_{j2}$, $\cdots$, $\hat{N}_{jn}$ are i.i.d. and $\hat{N}_{j1}\sim\mathcal{N}(0,\sigma_{j,N2}^2)$. In the following theorem, we prove that the average error probability $\mathbb{E}\{\tilde{\epsilon}_j\}$ under $\hat{\mathbf{n}}_j$ always lower bounds the actual error probability under $\mathbf{n}_j$, given a new lower bound on the average error probability $\hat{\epsilon}_{j1|\mathbf{n}_j}$ which is derived in Lemma~\ref{conv lower bounds}.

\begin{theorem}
\label{2nd domin conv}
    Given the definitions of $\mathbf{n}_j$ and $\hat{\mathbf{n}}_j$, if $\hat{\epsilon}_{j1|\mathbf{n}_j}$ is jointly convex in $\mathbf{n}_j$ and $\sigma_{j,N2}$ satisfies the conditions in Theorem~\ref{conv}, we have
    \begin{align}
            \int_{\mathcal{D}} \hat{\epsilon}_{j1|\mathbf{n}_j}(\mathbf{n}_j)f_{\mathbf{n}_j}(\mathbf{n}_j)\,\mathrm{d}\mathbf{n}_j \leq \int_{\mathcal{D}} \hat{\epsilon}_{j1|\mathbf{n}_j}(\mathbf{n}_j)f_{\hat{\mathbf{n}}_j}(\mathbf{n}_j)\,\mathrm{d}\mathbf{n}_j,\label{conv-vector-comparison}
    \end{align}
where $f_{\mathbf{n}_j}(\cdot)$ and $f_{\hat{\mathbf{n}}_j}(\cdot)$ are the PDFs of $\mathbf{n}_j$ and $\hat{\mathbf{n}}_j$, respectively, and $\mathcal{D}=\mathbb{R}^{n}$ represents the feasible region of the noise.

\end{theorem}
\noindent\textit{Proof:}
The initial part is analogous to the proof of Theorem~\ref{2nd domin} in Appendix~\ref{2nd domin proof}. Accordingly, it is equivalent for \eqref{conv-vector-comparison} to prove the following
\begin{align}
    \Delta^*\triangleq\int_{\mathcal{D}}\hat{\epsilon}_{j1|\mathbf{n}_j}^{(2)}(\mathbf{n})\left(S_{\mathbf{n}_j}(\mathbf{n})-S_{\hat{\mathbf{n}}_j}(\mathbf{n})\right)\,\dd \mathbf{n}\geq0.
\end{align}
By partitioning the integration domain according to the support of $S_{\hat{\mathbf{n}}_j}$, namely, $\hat{\mathcal{D}}=\{\mathbf{n}|S_{\hat{\mathbf{n}}_j}\neq0\}$ and $\mathcal{D}\setminus\hat{\mathcal{D}}$, we have
\begin{align}
    \Delta^*&=-\int_{\mathcal{D}\setminus\hat{\mathcal{D}}}\hat{\epsilon}_{j1|\mathbf{n}_j}^{(2)}(\mathbf{n})S_{\hat{\mathbf{n}}_j}(\mathbf{n})\,\dd \mathbf{n} +\int_{\hat{\mathcal{D}}}\hat{\epsilon}_{j1|\mathbf{n}_j}^{(2)}(\mathbf{n})\left(S_{\mathbf{n}_j}(\mathbf{n})-S_{\hat{\mathbf{n}}_j}(\mathbf{n})\right)\,\dd \mathbf{n}\\
    &=\int_{\hat{\mathcal{D}}}\hat{\epsilon}_{j1|\mathbf{n}_j}^{(2)}(\mathbf{n})\left(S_{\mathbf{n}_j}(\mathbf{n})-S_{\hat{\mathbf{n}}_j^*}(\mathbf{n})\right)\,\dd \mathbf{n}\label{delta1}\\
    &\quad-\int_{\hat{\mathcal{D}}}\hat{\epsilon}_{j1|\mathbf{n}_j}^{(2)}(\mathbf{n})S_{\delta}(\mathbf{n})\,\dd \mathbf{n} - \int_{\mathcal{D}\setminus\hat{\mathcal{D}}}\hat{\epsilon}_{j1|\mathbf{n}_j}^{(2)}(\mathbf{n})S_{\hat{\mathbf{n}}_j}(\mathbf{n})\,\dd \mathbf{n},\label{delta2}
\end{align}
where $S_{\delta}(\mathbf{n})\triangleq S_{\hat{\mathbf{n}}_j}(\mathbf{n})-S_{\hat{\mathbf{n}}_j^*}(\mathbf{n})$. Then, we characterize the maximum value of $\sigma_{j,N2}^*$ in \eqref{a2-sigma} such that \eqref{delta1} remains nonnegative. Details are provided in Appendix~\ref{pdelta1}. Given $\sigma_{j,N2}^*$, we are able to give the maximum value of $\sigma_{j,N2}$ such that \eqref{delta2} remains nonnegative. Equivalently, we need to derive the maximum value of $\sigma_{j,N2}$, such that
\begin{align}
    \int_{\mathcal{D}}\hat{\epsilon}_{j1|\mathbf{n}_j}^{(2)}(\mathbf{n})S_{\hat{\mathbf{n}}_j}(\mathbf{n})\,\dd \mathbf{n}\leq\int_{\hat{\mathcal{D}}}\hat{\epsilon}_{j1|\mathbf{n}_j}^{(2)}(\mathbf{n})S_{\hat{\mathbf{n}}_j^*}(\mathbf{n})\,\dd \mathbf{n},
\end{align}
as given in \eqref{conv-2} as the condition of Theorem~\ref{conv}.
$\hfill\square$

In order to employ Theorem~\ref{2nd domin conv} to prove Theorem~\ref{conv}, the remaining is to justify the condition of convexity of $\hat{\epsilon}_{j1|\mathbf{n}_j}$, which is done in the following subsection.

\subsection{Derivation of the Convex Lower Bound of $\mathbb{E}_{Y}\{\tilde{\epsilon}_{j1|\mathbf{n}_j}\}$}
According to \eqref{3-4} and the isotropic properties of the codewords and additive noise, the averaged lower bound derived below depends on $\mathbf{n}_j$ only through $W=\Vert\mathbf{n}_j\Vert$. However, convexity with respect to the scalar magnitude $W$ is \textit{not} in general equivalent to joint convexity with respect to $\mathbf{n}_j$. By Lemma~\ref{radial convex lemma}, a sufficient condition is that the radial function be both convex and non-decreasing in $W$. Therefore, in the following we start from \eqref{2-5} and \eqref{2-6}, derive a radial lower bound $\hat{\epsilon}_{j1|W}\triangleq\mathbb{E}_{Y}\{\hat{\epsilon}_{j1|\mathbf{n}_j}\}$, prove its convexity in $W$, and then verify its monotonicity before invoking the vector stochastic-dominance argument.

\begin{lemma}
    \label{conv lower bounds}
A convex lower bound on $\tilde{\epsilon}_{j1|W}$ is given by
\begin{align}
     \tilde{\epsilon}_{j1|W}\geq\hat{\epsilon}_{j1|W}\triangleq\hat{\epsilon}_{j1|W}^{11}+\hat{\epsilon}_{j1|W}^{12}+\hat{\epsilon}_{j1|W}^{13}+\hat{\epsilon}_{j1|W}^{21}+\hat{\epsilon}_{j1|W}^{22}+\hat{\epsilon}_{j1|W}^{23},\label{conv-1-proof}
\end{align}
where expressions for each term are given in \eqref{conv-101}-\eqref{conv-106}.
\end{lemma}

\noindent\textit{Proof:}
We start with the PDF of $Y$, i.e., $f_{Y|W}(y)$ as
\begin{align}
    f_{Y|W}(y|w)&=n(h_jP_j)^{-\frac{n}{2}}y^{n-1}\times\left\{\begin{array}{cl}
    1,&0\leq y\leq |\sqrt{h_jP_j}-w|\\[6pt]
     1-\frac{1}{2}I_{\sin^2\beta}\left(\frac{n-1}{2},\frac{1}{2}\right),&|\sqrt{h_jP_j}-w|<y\leq \sqrt{|h_jP_j-w^2|}\\[6pt]
     \frac{1}{2}I_{\sin^2\beta}\left(\frac{n-1}{2},\frac{1}{2}\right),& \sqrt{|h_jP_j-w^2|}<y\leq \sqrt{h_jP_j}+w,
     \end{array}\right.\label{2-4}
\end{align}
which is proved in Appendix~\ref{fYW}. Then, we derive the lower bounds of $\tilde{\epsilon}_{j1|W}$ for each case given in \eqref{2-5} and \eqref{2-6} respectively as
\begin{equation}
\begin{aligned}
    \tilde{\epsilon}_{j1|\mathbf{n}_j} \geq\left\{\begin{array}{cl}
        \left(\frac{w}{\sqrt{h_jP_j}}\right)^n, & 0\leq y\leq \sqrt{h_jP_j}-w\quad\quad\quad\quad\quad\,\text{(case 1)} \\
        \frac{1}{2}\left(\frac{w}{\sqrt{h_jP_j}}\right)^n, & \sqrt{h_jP_j}-w<y\leq \sqrt{h_jP_j-w^2}\text{ (case 2)} \\
        \frac{1}{2}I_{\sin^2\alpha}\left(\tfrac{n+1}{2},\tfrac{1}{2}\right), & \sqrt{h_jP_j-w^2}<y\leq \sqrt{h_jP_j}+w\text{ (case 3)},
    \end{array}\right.\label{2-5a}
    \end{aligned}
\end{equation}
for $w<\sqrt{h_jP_j}$, and
    \begin{equation}
\begin{aligned}
     \tilde{\epsilon}_{j1|\mathbf{n}_j}\geq\left\{\begin{array}{cl}
    1,&0\leq y\leq w-\sqrt{h_jP_j}\quad\quad\quad\quad\quad\,\text{(case 4)}\\
    \frac{1}{2},&w-\sqrt{h_jP_j}<y\leq \sqrt{w^2-h_jP_j}\text{ (case 5)}\\
     \frac{1}{2}I_{\sin^2\alpha}\left(\frac{n+1}{2},\frac{1}{2}\right),& \sqrt{w^2-h_jP_j}<y\leq \sqrt{h_jP_j}+w\text{ (case 6)},
     \end{array}\right.\label{2-6a}
\end{aligned}
\end{equation}
for $w\geq\sqrt{h_jP_j}$. Then by calculating
\begin{align}
    \tilde{\epsilon}_{j1|W}&\geq\int_{0}^{\sqrt{h_jP_j}+w}\tilde{\epsilon}_{j1|\mathbf{n}_j}f_{Y|W}(y|w)\,\dd y,
\end{align}
we obtain \eqref{conv-1-proof}, where $\hat{\epsilon}_{j1|W}^{11}$, $\hat{\epsilon}_{j1|W}^{12}$, and $\hat{\epsilon}_{j1|W}^{13}$ represent case 1, case 2, and case 3, respectively, while $\hat{\epsilon}_{j1|W}^{21}$, $\hat{\epsilon}_{j1|W}^{22}$, and $\hat{\epsilon}_{j1|W}^{23}$ represent case 4, case 5, and case 6, respectively. Proofs for $\hat{\epsilon}_{j1|W}^{13}$ and $\hat{\epsilon}_{j1|W}^{23}$ are provided in Appendix~\ref{ehat}, while proofs for $\hat{\epsilon}_{j1|W}^{11}$, $\hat{\epsilon}_{j1|W}^{12}$, $\hat{\epsilon}_{j1|W}^{21}$, and $\hat{\epsilon}_{j1|W}^{22}$ are relatively simple and therefore omitted for brevity.

Finally, we prove that $\hat{\epsilon}_{j1|W}$ is convex in $w$. Proofs for $\hat{\epsilon}_{j1|W}^{13}$ and $\hat{\epsilon}_{j1|W}^{23}$ are provided in Appendix~\ref{ehat convex}, while proofs for $\hat{\epsilon}_{j1|W}^{11}$, $\hat{\epsilon}_{j1|W}^{12}$, $\hat{\epsilon}_{j1|W}^{21}$, and $\hat{\epsilon}_{j1|W}^{22}$ are relatively simple and omitted for brevity. In addition, $\hat{\epsilon}_{j1|W}(w)\geq0$ and direct substitution gives $\hat{\epsilon}_{j1|W}(0)=0$. For arbitrary $0\leq w_1\leq w_2$, convexity therefore yields
\begin{align}
\hat{\epsilon}_{j1|W}(w_1)
&=\hat{\epsilon}_{j1|W}\left(\frac{w_1}{w_2}w_2+\left(1-\frac{w_1}{w_2}\right)0\right)\notag\\
&\leq\frac{w_1}{w_2}\hat{\epsilon}_{j1|W}(w_2)
\leq\hat{\epsilon}_{j1|W}(w_2),
\end{align}
where the case $w_2=0$ is trivial. Hence, $\hat{\epsilon}_{j1|W}$ is also non-decreasing. Lemma~\ref{radial convex lemma} then guarantees that $\hat{\epsilon}_{j1|\mathbf{n}_j}(\mathbf{n}_j)\triangleq\hat{\epsilon}_{j1|W}(\Vert\mathbf{n}_j\Vert)$ is jointly convex in the noise vector, which is the condition required in Theorem~\ref{2nd domin conv}.
$\hfill\square$

In conclusion, by constructing a convex lower bound $\hat{\epsilon}_{j1|W}$ and a new interference-plus-noise term $\hat{\mathbf{n}}_j$, and by invoking Theorem~\ref{2nd domin conv}, the proof for Theorem~\ref{conv} completes.

\section{Proof of Theorem~\ref{thm:jnn-two-user}}
\label{proof of jnn-two-user}
In this section, we derive an upper bound of the average error probability of each user under JNN decoding. Due to the symmetry of the random codebook construction, without loss of generality, suppose that the transmitted message pair is $(w_1,w_2)=(1,1)$ and denote $\mathbf u_1\triangleq\mathbf u_1(1)$ and $\mathbf u_2\triangleq\mathbf u_2(1)$, then we have
\begin{align}
    \mathbf y
    =
    \mathbf u_1+\mathbf u_2+\mathbf z.
    \label{eq:jnn-y-transmitted}
\end{align}

Under JNN decoding, an error occurs if at least one competing message pair has a Euclidean distance to $\mathbf y$ no larger than that of the transmitted message pair. All competing message pairs can be divided into three mutually exclusive classes according to which users are decoded incorrectly, under a two-user case, we have in total three error events
\begin{align}
    \mathcal W_1
    &\triangleq
    \left\{
        (\bar w_1,1):
        \bar w_1\neq1
    \right\},
    \\
    \mathcal W_2
    &\triangleq
    \left\{
        (1,\bar w_2):
        \bar w_2\neq1
    \right\},
    \\
    \mathcal W_{12}
    &\triangleq
    \left\{
        (\bar w_1,\bar w_2):
        \bar w_1\neq1,\,
        \bar w_2\neq1
    \right\}.
    \label{eq:jnn-three-classes}
\end{align}
The corresponding numbers of competing message pairs are $M_1-1$, $M_2-1$, and $(M_1-1)(M_2-1)$, respectively.

\subsection{Error Events of JNN}
We first consider a competing pair $(\bar w_1,1)\in\mathcal W_1$. This competitor is chosen as the transmitted pair by the JNN decoder if
\begin{align}
    \left\|
        \mathbf y
        -
        \mathbf u_1(\bar w_1)
        -
        \mathbf u_2
    \right\|^2
    \leq
    \left\|
        \mathbf y
        -
        \mathbf u_1
        -
        \mathbf u_2
    \right\|^2.
    \label{eq:jnn-class1-metric}
\end{align}
Substituting \eqref{eq:jnn-y-transmitted} into \eqref{eq:jnn-class1-metric} gives
\begin{align}
    \left\|
        \mathbf z
        +
        \mathbf u_1
        -
        \mathbf u_1(\bar w_1)
    \right\|^2
    \leq
    \|\mathbf z\|^2.
    \label{eq:jnn-class1-event}
\end{align}
Therefore, conditioned on $(\mathbf u_1,\mathbf z)$, the probability of the pairwise event in \eqref{eq:jnn-class1-event} is exactly $p_1(\mathbf u_1,\mathbf z)$ defined in \eqref{eq:jnn-p-condition}. Notice that $\mathbf u_2$ disappears from \eqref{eq:jnn-class1-event}; hence, when only user 1 is incorrectly decoded, the codeword of user 2 does not act as interference under joint decoding.

Analogously, for a competing pair $(1,\bar w_2)\in\mathcal W_2$, the pairwise error event is
\begin{align}
    \left\|
        \mathbf z
        +
        \mathbf u_2
        -
        \mathbf u_2(\bar w_2)
    \right\|^2
    \leq
    \|\mathbf z\|^2,
    \label{eq:jnn-class2-event}
\end{align}
whose conditional probability is $p_2(\mathbf u_2,\mathbf z)$.

Finally, consider $(\bar w_1,\bar w_2)\in\mathcal W_{12}$. The corresponding JNN metric satisfies
\begin{align}
    &\left\|
        \mathbf y
        -
        \mathbf u_1(\bar w_1)
        -
        \mathbf u_2(\bar w_2)
    \right\|^2
    \leq
    \left\|
        \mathbf y
        -
        \mathbf u_1
        -
        \mathbf u_2
    \right\|^2.
    \label{eq:jnn-class12-event}
\end{align}
Conditioned on $(\mathbf u_1,\mathbf u_2,\mathbf z)$, its probability is $p_{12}(\mathbf u_1,\mathbf u_2,\mathbf z)$.

Let $\mathcal E_1(\bar w_1)$, $\mathcal E_2(\bar w_2)$, and $\mathcal E_{12}(\bar w_1,\bar w_2)$ denote the events in \eqref{eq:jnn-class1-event}, \eqref{eq:jnn-class2-event}, and \eqref{eq:jnn-class12-event}, respectively. Applying the union bound conditioned on $(\mathbf u_1,\mathbf u_2,\mathbf z)$ yields
\begin{align}
    \Pr
    \left\{
        \mathcal E_{\mathrm J}
        \mid
        \mathbf u_1,\mathbf u_2,\mathbf z
    \right\}\leq
    \min
    \bigg\{
        1,\,
        (M_1-1)p_1
        +(M_2-1)p_2
        +(M_1-1)(M_2-1)p_{12}
    \bigg\}.
    \label{eq:jnn-union-min}
\end{align}
Averaging \eqref{eq:jnn-union-min} over the transmitted VURC codewords and the Gaussian noise proves \eqref{eq:jnn-rcu}.

\subsection{Explicit Error Probability for Each Event}
We next derive the explicit expressions in \eqref{eq:jnn-pi}--\eqref{eq:jnn-p12}. Consider an arbitrary pairwise
error event and define the displacement between the transmitted and competing superposed codewords as $\boldsymbol\delta$. The pairwise error event can always be written as
\begin{align}
    \|\mathbf z+\boldsymbol\delta\|^2
    \leq
    \|\mathbf z\|^2.
    \label{eq:jnn-generic-pep}
\end{align}
Expanding the squared norm on the left-hand side gives
\begin{align}
    \mathbf z^{\mathsf T}\boldsymbol\delta
    \leq
    -\frac{1}{2}
    \|\boldsymbol\delta\|^2.
    \label{eq:jnn-pep-inner}
\end{align}
Conditioned on $\boldsymbol\delta$, we have
\begin{align}
    \mathbf z^{\mathsf T}\boldsymbol\delta
    \sim
    \mathcal N
    \left(
        0,
        \frac{\sigma^2}{n}
        \|\boldsymbol\delta\|^2
    \right).
    \label{eq:jnn-inner-gaussian}
\end{align}
For $\boldsymbol\delta\neq\mathbf0$, standardizing \eqref{eq:jnn-pep-inner} therefore gives
\begin{align}
    \Pr
    \left\{
        \|\mathbf z+\boldsymbol\delta\|^2
        \leq
        \|\mathbf z\|^2
        \,\middle|\,
        \boldsymbol\delta
    \right\}
    =
    Q
    \left(
        \frac{\sqrt n
        \|\boldsymbol\delta\|}
        {2\sigma}
    \right).
    \label{eq:jnn-pep}
\end{align}
Since the VURC ensemble is continuous, we have $\Pr\{\boldsymbol\delta=\mathbf0\}=0$, and hence \eqref{eq:jnn-pep} determines the ensemble-averaged pairwise error probability.

For the first error class, i.e., $\mathcal W_1$ and $\mathcal W_2$, define $\boldsymbol\Delta_i\triangleq\mathbf U_i-\bar{\mathbf U}_i$ and $D_i\triangleq\|\boldsymbol\Delta_i\|$, where $\mathbf U_i$ and $\bar{\mathbf U}_i$ are independent and uniformly distributed inside $\mathcal B_n(a_i)$. We first derive the PDF of
$D_i$. Since
\begin{align}
    f_{\mathbf U_i}(\mathbf u)
    =
    \frac{
        \mathbf 1
        \{\|\mathbf u\|\leq a_i\}
    }{
        V_n(a_i)
    },
\end{align}
the PDF of $\boldsymbol\Delta_i=\mathbf U_i-\bar{\mathbf U}_i$ is the convolution
\begin{align}
    f_{\boldsymbol\Delta_i}(\boldsymbol\delta)
    &=
    \int_{\mathbb R^n}
    f_{\mathbf U_i}(\mathbf u)
    f_{\bar{\mathbf U}_i}
    (\mathbf u-\boldsymbol\delta)
    \,\mathrm d\mathbf u
    \\
    &=
    \frac{
        V_{\mathrm{ov}}
        \left(
            a_i,\|\boldsymbol\delta\|
        \right)
    }{
        V_n^2(a_i)
    },
    \label{eq:jnn-difference-vector-pdf}
\end{align}
where $V_{\mathrm{ov}}(a,d)$ is the intersection volume of two $n$-dimensional hyperspheres with the same radius $a$ and center separation $d$. For $0\leq d\leq2a$, the intersection consists of two identical spherical caps and we have
\begin{align}
    V_{\mathrm{ov}}(a,d)
    =
    V_n(a)
    I_{1-\frac{d^2}{4a^2}}
    \left(
        \frac{n+1}{2},
        \frac12
    \right).
    \label{eq:jnn-overlap}
\end{align}
Because $f_{\boldsymbol\Delta_i}(\boldsymbol\delta)$ depends only on $\|\boldsymbol\delta\|$, $\boldsymbol\Delta_i$ is spherically symmetric. Consequently, the PDF of its magnitude satisfy
\begin{align}
    f_{D_i}(d)
    &=
    S_n(d)
    f_{\boldsymbol\Delta_i}(d)
    \\
    &=
    \frac{nd^{n-1}}{a_i^n}
    I_{1-\frac{d^2}{4a_i^2}}
    \left(
        \frac{n+1}{2},
        \frac12
    \right),
    \label{eq:jnn-distance-pdf-proof}
\end{align}
where $S_n(d)=2\pi^{n/2}d^{n-1}/\Gamma(n/2)$ denotes the surface area of an $n$-dimensional sphere of radius $d$, \eqref{eq:jnn-distance-pdf-proof} follows from $S_n(d)/V_n(a_i)=nd^{n-1}/a_i^n$. This proves \eqref{eq:jnn-distance-pdf}.

For the event in which only user $i$ is incorrectly decoded, the displacement in \eqref{eq:jnn-pep} is simply
$\boldsymbol\delta=\boldsymbol\Delta_i$. Averaging \eqref{eq:jnn-pep} over $D_i$ therefore yields
\begin{samepage}
\begin{align}
    \bar p_i
    &\triangleq
    \mathbb E
    \left\{
        p_i
    \right\}\\
    &=
    \mathbb E_{D_i}
    \left[
        Q
        \left(
            \frac{\sqrt nD_i}{2\sigma}
        \right)
    \right]\\
    &=
    \int_0^{2a_i}
    Q
    \left(
        \frac{\sqrt n\,d}{2\sigma}
    \right)
    f_{D_i}(d)\,\mathrm dd,
    \label{eq:jnn-pi-proof}
\end{align}
\end{samepage}
which proves \eqref{eq:jnn-pi}.

Then, we derive the pairwise error probability associated with $\mathcal W_{12}$. In this case, from \eqref{eq:jnn-class12-event}, the displacement is $\boldsymbol\Delta_{12}=\boldsymbol\Delta_1+\boldsymbol\Delta_2$. The two difference vectors are independent because the two users generate their codebooks independently. 
Conditioned on $D_1=d_1$ and $D_2=d_2$, and let $T$ denote the cosine of the angle between
$\boldsymbol\Delta_1$ and $\boldsymbol\Delta_2$, we have
\begin{align}
    \|\boldsymbol\Delta_{12}\|^2
    =
    d_1^2+d_2^2+2d_1d_2T.
    \label{eq:jnn-joint-distance}
\end{align}
Since the directions of two independent spherically symmetric random vectors are independently and uniformly distributed over $\mathbb S^{n-1}$, the random variable $T$ has the PDF \cite[eq. (44)]{TIT}
\begin{align}
    f_T(t)
    =
    \frac{\Gamma(n/2)}
    {\sqrt{\pi}\Gamma((n-1)/2)}
    (1-t^2)^{\frac{n-3}{2}},
    \qquad
    -1<t<1.
\end{align}
Conditioning first on $(D_1,D_2,T)$ and applying \eqref{eq:jnn-pep}, we obtain
\begin{align}
    &\Pr
    \left\{
        \mathcal E_{12}
        \mid
        D_1=d_1,D_2=d_2,T=t
    \right\}
    =
    Q
    \left(
        \frac{\sqrt n}{2\sigma}
        \sqrt{
            d_1^2+d_2^2+2d_1d_2t
        }
    \right).
    \label{eq:jnn-p12-conditioned}
\end{align}
Since $D_1$, $D_2$, and the two directions are mutually independent, averaging \eqref{eq:jnn-p12-conditioned} over $d_1$, $d_2$, and $t$ gives
\begin{align}
    \bar p_{12}
    =&
    \int_0^{2a_1}
    \int_0^{2a_2}
    \int_{-1}^{1}
    Q
    \left(
        \frac{\sqrt n}{2\sigma}
        \sqrt{
            d_1^2+d_2^2+2d_1d_2t
        }
    \right)
    \nonumber\\
    &\qquad\times
    f_{D_1}(d_1)
    f_{D_2}(d_2)
    f_T(t)
    \,\mathrm dt\,\mathrm dd_2\,\mathrm dd_1,
\end{align}
which is exactly \eqref{eq:jnn-p12}.

It remains to obtain \eqref{eq:jnn-explicit-bound}. From \eqref{eq:jnn-rcu}, define
\begin{align}
    Z_{\mathrm J}
    \triangleq
    (M_1-1)p_1
    +(M_2-1)p_2
    +(M_1-1)(M_2-1)p_{12},
\end{align}
we have
\begin{align}
    \mathbb E
    \left[
        \min\{1,Z_{\mathrm J}\}
    \right]
    &\leq
    \min
    \left\{
        1,\,
        \mathbb E\{Z_{\mathrm J}\}
    \right\}\\
    &=
    \min
    \bigg\{
        1,\,
        (M_1-1)\bar p_1
        +(M_2-1)\bar p_2+(M_1-1)(M_2-1)\bar p_{12}
    \bigg\},
\end{align}
which proves \eqref{eq:jnn-explicit-bound}. This completes the proof.

\section{Proof of Corollary~\ref{cor:jnn-closed-form}}
\label{proof of jnn-closed-form}
We first derive closed-form upper bounds for the two single-user error events, i.e., $\bar p_1$ and $\bar p_2$. Denote the magnitude of the Gaussian noise by $W\triangleq\|\mathbf z\|$. Conditioned on the transmitted effective codeword $\mathbf u_i$ and the noise $\mathbf z$, according to \eqref{eq:jnn-p-condition}, the pairwise error event corresponding to $\mathcal W_i$ is
\begin{align}
    \left\|
        \mathbf z+\mathbf u_i-\bar{\mathbf U}_i
    \right\|
    \leq W.
    \label{eq:jnn-volume-event-i}
\end{align}
Equivalently, the competing codeword $\bar{\mathbf U}_i$ needs to fall inside the $n$-dimensional hypersphere $\mathcal B_n(\mathbf u_i+\mathbf z,W)$. Since
$\bar{\mathbf U}_i$ is uniformly distributed inside $\mathcal B_n(\sqrt{h_iP_i})$, the conditional
probability can be written as
\begin{align}
    p_i(\mathbf u_i,\mathbf z)
    &=
    \frac{
        V\left[
            \mathcal B_n(\sqrt{h_iP_i})
            \cap
            \mathcal B_n(\mathbf u_i+\mathbf z,W)
        \right]
    }{
        V_n(\sqrt{h_iP_i})
    }\\
    &\leq
    \min\left\{
        1,\,
        \frac{V_n(W)}{V_n(\sqrt{h_iP_i})}
    \right\}\label{eq:jnn-pi-volume-bound-1}\\
   &=
    \min\left\{
        1,\,
        \left(\frac{W}{\sqrt{h_iP_i}}\right)^n
    \right\},
    \label{eq:jnn-pi-volume-bound}
\end{align}
where \eqref{eq:jnn-pi-volume-bound-1} follows since the intersection volume cannot exceed either
$V_n(W)$ or $V_n(\sqrt{h_iP_i})$.

We then average \eqref{eq:jnn-pi-volume-bound} over the transmitted codeword and noise. Since its right-hand side only depends on $W$, we have
\begin{align}
    \bar p_i
    =
    \mathbb E\{p_i\}
    \leq
    \mathbb E_W
    \left\{
        \min\left[
            1,\left(\frac{W}{\sqrt{h_iP_i}}\right)^n
        \right]
    \right\}.
    \label{eq:jnn-pi-W}
\end{align}
Since the elements of $\mathbf z$ are i.i.d. Gaussian random variables with variance $\sigma^2/n$. Therefore,
$\sqrt n W/\sigma$ follows a $\chi$-distribution with $n$ degrees of freedom, and the PDF of $W$ follows
\begin{align}
    f_W(w)
    =
    \frac{
        2^{1-\frac n2}n^{\frac n2}
    }{
        \Gamma\left(\frac n2\right)\sigma^n
    }
    w^{n-1}
    \exp\left(
        -\frac{nw^2}{2\sigma^2}
    \right).
    \label{eq:jnn-W-pdf}
\end{align}
According to \eqref{eq:jnn-pi-W}, we partition the integration region at $w=\sqrt{h_iP_i}$ and obtain
\begin{align}
    \bar p_i
    \leq&
    \int_0^{\sqrt{h_iP_i}}
    \left(\frac{w}{\sqrt{h_iP_i}}\right)^n
    f_W(w)\,\mathrm dw
    +
    \int_{\sqrt{h_iP_i}}^{\infty}
    f_W(w)\,\mathrm dw.
    \label{eq:jnn-pi-split}
\end{align}
For the first term in \eqref{eq:jnn-pi-split}, substituting $x=nw^2/(2\sigma^2)$ gives
\begin{samepage}
\begin{align}
    \int_0^{\sqrt{h_iP_i}}
    \left(\frac{w}{\sqrt{h_iP_i}}\right)^n
    f_W(w)\,\mathrm dw
    &=
    \frac{
        2^{1-\frac n2}n^{\frac n2}
    }{
        \Gamma\left(\frac n2\right)
        \sigma^n(h_iP_i)^{n/2}
    }
    \int_0^{\sqrt{h_iP_i}}
    w^{2n-1}
    \exp\left(
        -\frac{nw^2}{2\sigma^2}
    \right)
    \mathrm dw
    \\
    &=
    \left(
        \frac{2\sigma^2}{nh_iP_i}
    \right)^{\frac n2}
    \frac{
        \gamma\left(
            n,\frac{nh_iP_i}{2\sigma^2}
        \right)
    }{
        \Gamma\left(\frac n2\right)
    }
    \\
    &=
    \left(
        \frac{n\gamma_i}{2}
    \right)^{-\frac n2}
    \frac{
        \gamma\left(
            n,\frac{n\gamma_i}{2}
        \right)
    }{
        \Gamma\left(\frac n2\right)
    },
    \label{eq:jnn-pi-first}
\end{align}
\end{samepage}
where $\gamma_i=h_iP_i/\sigma^2$. Similarly, the second term in \eqref{eq:jnn-pi-split} gives
\begin{align}
    \int_{\sqrt{h_iP_i}}^{\infty}
    f_W(w)\,\mathrm dw
    =
    \frac{
        \Gamma\left(
            \frac n2,\frac{nh_iP_i}{2\sigma^2}
        \right)
    }{
        \Gamma\left(\frac n2\right)
    }
    =
    \frac{
        \Gamma\left(
            \frac n2,\frac{n\gamma_i}{2}
        \right)
    }{
        \Gamma\left(\frac n2\right)
    }.
    \label{eq:jnn-pi-second}
\end{align}
Combining \eqref{eq:jnn-pi-first} and \eqref{eq:jnn-pi-second}, we obtain $\bar p_i\leq\mathcal A_n(\gamma_i), i=1,2$.

We then derive a closed-form upper bound for the joint error event $\bar p_{12}$. Define the superposition of two independently generated competing codewords as $\bar{\mathbf S}\triangleq\bar{\mathbf U}_1+\bar{\mathbf U}_2$. Since $\bar{\mathbf U}_1$ and $\bar{\mathbf U}_2$ are independently and uniformly distributed inside $\mathcal B_n(\sqrt{h_1P_1})$ and $\mathcal B_n(\sqrt{h_2P_2})$, respectively, their PDFs are
\begin{align}
    f_{\bar{\mathbf U}_i}(\mathbf u)
    =
    \frac{
        \mathbf 1\{\|\mathbf u\|\leq \sqrt{h_iP_i}\}
    }{
        V_n(\sqrt{h_iP_i})
    },
    \qquad i=1,2.
    \label{eq:jnn-competitor-pdf}
\end{align}
Accordingly, the PDF of $\bar{\mathbf S}$ is given by the convolution
\begin{align}
    f_{\bar{\mathbf S}}(\mathbf s)
    &=
    \int_{\mathbb R^n}
    f_{\bar{\mathbf U}_1}(\mathbf u)
    f_{\bar{\mathbf U}_2}(\mathbf s-\mathbf u)
    \,\mathrm d\mathbf u
    \\
    &=
    \frac{
        V\left[
            \mathcal B_n(\sqrt{h_1P_1})
            \cap
            \mathcal B_n(\mathbf s,\sqrt{h_2P_2})
        \right]
    }{
        V_n(\sqrt{h_1P_1})V_n(\sqrt{h_2P_2})
    },
    \label{eq:jnn-sum-pdf}
\end{align}
where $\mathcal B_n(\mathbf s,\sqrt{h_2P_2})$ denotes an $n$-dimensional hypersphere centered at $\mathbf s$ with radius $\sqrt{h_2P_2}$. The intersection volume in
\eqref{eq:jnn-sum-pdf} satisfies
\begin{align}
    V\left[
        \mathcal B_n(\sqrt{h_1P_1})
        \cap
        \mathcal B_n(\mathbf s,\sqrt{h_2P_2})
    \right]
    \leq
    \min\left\{
        V_n(\sqrt{h_1P_1}),V_n(\sqrt{h_2P_2})
    \right\}.
    \label{eq:jnn-sum-volume}
\end{align}
Thus, letting $a_{12}\triangleq\max\{\sqrt{h_1P_1},\sqrt{h_2P_2}\}$, for $\forall\mathbf s\in\mathbb R^n$, we have
\begin{align}
    f_{\bar{\mathbf S}}(\mathbf s)
    &\leq
    \frac{
        \min\{V_n(\sqrt{h_1P_1}),V_n(\sqrt{h_2P_2})\}
    }{
        V_n(\sqrt{h_1P_1})V_n(\sqrt{h_2P_2})
    }
    \\
    &=
    \frac{1}{
        \max\{V_n(\sqrt{h_1P_1}),V_n(\sqrt{h_2P_2})\}
    }\\
    &=
    \frac{1}{V_n(a_{12})}.
    \label{eq:jnn-sum-density-bound}
\end{align}

According to the definition of $p_{12}$ in \eqref{eq:jnn-p-condition}, conditioned on $(\mathbf u_1,\mathbf u_2,\mathbf z)$, a joint pairwise
error occurs when $\bar{\mathbf S}\in\mathcal B_n(\mathbf u_1+\mathbf u_2+\mathbf z,W)$. Therefore, using \eqref{eq:jnn-sum-density-bound}, we have
\begin{align}
    p_{12}(\mathbf u_1,\mathbf u_2,\mathbf z)
    &=
    \int_{
        \mathcal B_n(
            \mathbf u_1+\mathbf u_2+\mathbf z,W
        )
    }
    f_{\bar{\mathbf S}}(\mathbf s)
    \,\mathrm d\mathbf s\\
    &\leq
    \frac{V_n(W)}{V_n(a_{12})}\\
    &\leq
    \min\left\{
        1,\,
        \left(\frac{W}{a_{12}}\right)^n
    \right\}.
    \label{eq:jnn-p12-volume-1}
\end{align}
Averaging \eqref{eq:jnn-p12-volume-1} over the transmitted codewords and Gaussian noise, following exactly the same derivation as \eqref{eq:jnn-pi-split}--\eqref{eq:jnn-pi-second}, we obtain $\bar p_{12}\leq\mathcal A_n(\gamma_{12})$,
where
\begin{align}
    \gamma_{12}
    =
    \frac{a_{12}^2}{\sigma^2}
    =
    \frac{\max\{h_1P_1,h_2P_2\}}{\sigma^2}
    =
    \max\{\gamma_1,\gamma_2\}.
\end{align}
This completes the proof.

\section{Proof of Theorem~\ref{2nd domin}}
\label{2nd domin proof}
For user $i$, since $\mathbf{x}_i$ is uniformly distributed over the $n$-dimensional hypersphere of radius $\sqrt{P_i}$,
$\sqrt{h_i}\mathbf{x}_i$ is uniformly distributed over the $n$-dimensional hypersphere of radius $\sqrt{h_iP_i}$. It admits the
spherical representation
\begin{align}
    \sqrt{h_i}\mathbf{x}_i
    =R_i\mathbf{U}_i,
    \label{spherical-vurc}
\end{align}
where $\mathbf{U}_i$ is uniformly distributed over the unit sphere, $R_i$ is independent of $\mathbf{U}_i$, and
\begin{align}
    f_{R_i}(r)
    =
    \frac{nr^{n-1}}{(h_iP_i)^{n/2}},
    \qquad
    0\leq r\leq\sqrt{h_iP_i}.
    \label{radius-vurc}
\end{align}
Construct $\mathbf{q}_i\sim\mathcal{N}\left(\mathbf{0},\frac{h_iP_i}{n}\mathbf{I}_n\right)$, the Gaussian vector is also isotropic and can be written as
$\mathbf{q}_i=G_i\mathbf{V}_i$, where $\mathbf{V}_i$ is uniform over the unit sphere and independent of $G_i$. The PDF of its radius is
\begin{align}
    f_{G_i}(r)
    =
    \frac{2^{1-\frac n2}}{\Gamma(\frac n2)}
    \left(
        \frac{n}{h_iP_i}
    \right)^{\frac n2}
    r^{n-1}
    \exp\left(
        -\frac{nr^2}{2h_iP_i}
    \right).
    \label{radius-gaussian}
\end{align}
We first establish the convex order $R_i\preceq_{\rm icx}G_i$. By the stop-loss characterization of the convex order
\cite[Sec.~4.A]{StochasticOrders}, it is sufficient to prove
\begin{align}
    D_i(t)
    \triangleq
    \mathbb{E}\{(G_i-t)^+\}
    -
    \mathbb{E}\{(R_i-t)^+\}
    \geq0,
    \qquad t\geq0.
    \label{stoploss-D}
\end{align}
For $t\geq\sqrt{h_iP_i}$, the second term in \eqref{stoploss-D} is zero and therefore $D_i(t)\geq0$. It remains to consider
$0\leq t\leq\sqrt{h_iP_i}$. On this interval,
\begin{align}
    D_i^{(1)}(t)
    &=
    F_{G_i}(t)-F_{R_i}(t),\\
    D_i^{(2)}(t)
    &=
    f_{G_i}(t)-f_{R_i}(t).
\end{align}
From \eqref{radius-vurc} and \eqref{radius-gaussian}, we have
\begin{align}
    \frac{f_{G_i}(t)}{f_{R_i}(t)}
    =
    \frac{
        2^{1-\frac n2}n^{\frac n2-1}
    }{
        \Gamma(\frac n2)
    }
    \exp\left(
        -\frac{nt^2}{2h_iP_i}
    \right),
    \label{density-ratio}
\end{align}
which is strictly decreasing in $t$. Hence, $D_i^{(2)}(t)$ changes sign at most once, and if a sign change occurs it is from positive to negative. Consequently, $D_i^{(1)}(t)$ is first non-decreasing and then non-increasing. Moreover,
\begin{align}
    D_i^{(1)}(0)
    &=0,\\
    D_i^{(1)}\left(\sqrt{h_iP_i}\right)
    &=
    F_{G_i}\left(\sqrt{h_iP_i}\right)-1
    <0.
\end{align}
Thus, $D_i(t)$ has no interior local minimum on $[0,\sqrt{h_iP_i}]$; its minimum over this interval must occur at an endpoint. At $t=\sqrt{h_iP_i}$,
\begin{align}
    D_i\left(\sqrt{h_iP_i}\right)
    =
    \mathbb{E}
    \left\{
        \left(
            G_i-\sqrt{h_iP_i}
        \right)^+
    \right\}
    >0.
\end{align}
At $t=0$, we have
\begin{align}
    D_i(0)
    &=
    \mathbb{E}\{G_i\}
    -
    \mathbb{E}\{R_i\}
    \label{D0-1}\\
    &=
    \sqrt{h_iP_i}
    \left[
        \sqrt{\frac{2}{n}}
        \frac{
            \Gamma\left(\frac{n+1}{2}\right)
        }{
            \Gamma\left(\frac n2\right)
        }
        -
        \frac{n}{n+1}
    \right]
    \label{D0-2}\\
    &\geq
    \sqrt{h_iP_i}
    \left[
        \sqrt{\frac{n}{n+1}}
        -
        \frac{n}{n+1}
    \right]
    >0,
    \label{D0}
\end{align}
where \eqref{D0-2} follows from \eqref{radius-vurc} and the mean of a $\chi_n$ random variable, \eqref{D0} follows from Wendel's inequality
\begin{align}
    \left(\frac{x}{x+s}\right)^{1-s}
    \leq
    \frac{\Gamma(x+s)}{x^s\Gamma(x)},
    \qquad x>0,\quad 0<s<1,
\end{align}
by taking $x=n/2$ and $s=1/2$. Hence, $D_i(0)>0$. Together with $D_i(\sqrt{h_iP_i})>0$ and the absence of an interior local minimum, this proves \eqref{stoploss-D}, and therefore $R_i\preceq_{\rm icx}G_i$.

We next lift the radial order to the full random vectors. Let $\phi:\mathbb{R}^n\to\mathbb{R}$ be an arbitrary convex function and define its spherical average as $\bar{\phi}(r)\triangleq\mathbb{E}_{\mathbf{U}}\{\phi(r\mathbf{U})\}$, where $\mathbf{U}$ is uniform over the unit sphere. For every fixed
$\mathbf{u}$, $r\mapsto\phi(r\mathbf{u})$ is convex, so $\bar{\phi}(r)$ is convex. Since $\bar{\phi}(r)$ is even and every even convex function is non-decreasing on $[0,\infty)$; therefore, by $R_i\preceq_{\rm icx}G_i$, we have
\begin{align}
    \mathbb{E}
    \left\{
        \phi\left(\sqrt{h_i}\mathbf{x}_i\right)
    \right\}
    &=
    \mathbb{E}\{\bar{\phi}(R_i)\}\\
    &\leq
    \mathbb{E}\{\bar{\phi}(G_i)\}\\
    &=
    \mathbb{E}\{\phi(\mathbf{q}_i)\}.
\end{align}
Consequently, we have $\sqrt{h_i}\mathbf{x}_i\preceq_{\rm cx}\mathbf{q}_i$.

Finally, the multivariate convex order is preserved under addition of an independent random vector. Indeed, if $\mathbf{X}\preceq_{\rm cx}\mathbf{Y}$ and $\mathbf{Z}$ is independent of both, then
\begin{align}
    \mathbb{E}\{\phi(\mathbf{X}+\mathbf{Z})\}
    &=
    \mathbb{E}\{g(\mathbf{X})\}\\
    &\leq
    \mathbb{E}\{g(\mathbf{Y})\}\\
    &=
    \mathbb{E}\{\phi(\mathbf{Y}+\mathbf{Z})\}.
\end{align}
Applying this argument successively to the independent user codewords in \eqref{3-4} yields
\begin{align}
    \mathbf{n}_j
    \preceq_{\rm cx}
    \mathbf{n}_{j,G}
    \triangleq
    \sum_{m=j+1}^{K}\mathbf{q}_m+\mathbf{z},
    \label{sum-cx}
\end{align}
where
\begin{align}
    \mathbf{n}_{j,G}
    \sim
    \mathcal{N}
    \left(
        \mathbf{0},
        \frac{
            \sigma^2+\sum_{m=j+1}^{K}h_mP_m
        }{n}
        \mathbf{I}_n
    \right).
\end{align}
If $\sigma_{j,N1}^2$ is strictly larger than the variance in \eqref{sigmaN1-cx}, $\tilde{\mathbf{n}}_j$ can be represented as $\mathbf{n}_{j,G}+\mathbf{v}$ with an independent zero-mean Gaussian vector $\mathbf{v}$. Jensen's inequality then gives $\mathbf{n}_{j,G}\preceq_{\rm cx}\tilde{\mathbf{n}}_j$. Combining this relation with \eqref{sum-cx} proves \eqref{vector-cx-noise} and \eqref{ach-vector-cx}. This completes the proof.

\section{Derivation of $\sigma_{j,N2}^*$ Satisfying $\eqref{delta1}\geq0$}
\label{pdelta1}
Analogous to the derivation in Theorem~\ref{2nd domin}, we have
\begin{align}
    S_{\mathbf{n}_j}(\mathbf{n})&=\prod_{l=1}^{n}S_{N_{jl}}(N_{jl}),\\
    S_{\hat{\mathbf{n}}_j^*}(\mathbf{n})&=\prod_{l=1}^{n}S_{\hat{N}_{jl}^*}({N}_{jl}),
\end{align}
for any $l$, $S_{N_{jl}}(\cdot)$ are identical, and the same holds for $S_{\hat{N}_{jl}^*}(\cdot)$. Consequently, we only need to prove that for any $x\in\mathbb{R}$, $S_{N_{jl}}(x)-S_{\hat{N}_{jl}^*}(x)\geq0$ holds. 

Construct a new random variable $\hat{\mathbf{q}}_i=[\hat{Q}_{i1},\hat{Q}_{i2},\cdots,\hat{Q}_{in}]$, $i=1,2,\dots,K$ to replace $\mathbf{x}_i$, where $\hat{Q}_{i1}\sim \mathcal{N}\left[0,\left(\sigma_{j1,N2}^*\right)^2\right]$ and $\hat{Q}_{i1},\cdots,\hat{Q}_{in}$ are i.i.d.. Then, we define
\begin{align}
    \hat{\mathbf{n}}_j\triangleq\sum_{m=j+1}^{K}\hat{\mathbf{q}}_m+\mathbf{w}.
\end{align}
Together with \eqref{3-4}, as stochastic dominance is preserved with convolution, without loss of generality, we only need to prove that $X_{i1}\leq_{\text{SSD}}\hat{Q}_{i1}$, which is equivalent to prove
\begin{align}
    S_{X_{i1}}(x)-S_{\hat{Q}_{i1}}(x)\geq0,\,\,\forall{x\in\mathbb{R}},
\end{align}
where $S_{X_{i1}}=\int_{-\infty}^{x}\int_{-\infty}^{x'}f_{X_{i1}}(x'')\,\mathrm{d}x''\,\mathrm{d}x'$, $S_{\hat{Q}_{i1}}=\int_{-\infty}^{x}\int_{-\infty}^{x'}f_{\hat{Q}_{i1}}(x'')\,\dd  x'' \,\dd x'$, and 
\begin{align}
    f_{\hat{Q}_{i1}}(x)=\frac{1}{\sqrt{2\pi}\sigma_{j1,N2}^*}\exp{\left[-\frac{x^2}{2\left(\sigma_{j1,N2}^*\right)^2}\right]}. 
\end{align}
Define $G_*(x)\triangleq S_{X_{i1}}(x)-S_{\hat{Q}_{i1}}(x)$, our final goal is to prove that $G_*(x)\geq0\,\,\forall x\in\mathbb{R}$.

Analogous to the proof in Appendix~\ref{2nd domin proof}, $G_*(x)$ is an even function, therefore we only need to prove $G_*(x)\geq0\,\,\forall x\in\mathbb{R_+}$. Since $G_*^{(1)}(0)>0$ and $\lim_{x\to\infty}G_*^{(1)}(x)=0$, we only need to prove: 1) $G_*^{(2)}(0)>0$; 2) $G_*^{(2)}(x)$ has and only has one zero point within $x\in\mathbb{R}_+$. 

According to the definitions, we have
\begin{align}
    G_*^{(2)}(0)&=\frac{1}{\sqrt{2\pi}\sigma_{j1,N2}^*}-(h_iP_i)^{-\frac{1}{2}}B^{-1}\left(\frac{n+1}{2},\frac{1}{2}\right)\\
    &=\frac{1}{\sqrt{2\pi}\sigma_{j1,N2}^*}-\frac{1}{\sqrt{\pi}}(h_iP_i)^{-\frac{1}{2}}\frac{\Gamma(\frac{n}{2}+1)}{\Gamma(\frac{n+1}{2})}\label{pd-1}\\
    &\geq\frac{1}{\sqrt{2\pi}\sigma_{j1,N2}^*}-\frac{1}{\sqrt{\pi}}(h_iP_i)^{-\frac{1}{2}}\sqrt{\frac{n+1}{2}}\label{pd-2},
\end{align}
where \eqref{pd-1} follows that $B(a,b)=\frac{\Gamma(a)\Gamma(b)}{\Gamma(a+b)}$ and $\Gamma(\frac{1}{2})=\sqrt{\pi}$, \eqref{pd-2} employs Gautschi's inequality, which states that $x^{1-s}<\frac{\Gamma(x+1)}{\Gamma(x+s)}<(x+1)^{1-s}$. By forcing $G_*^{(2)}(0)>0$, we have
\begin{align}
    \sigma_{j1,N2}^*\leq\sqrt{\frac{h_iP_i}{n+1}}.\label{a2-sigma}
\end{align}

The proof that $G_*^{(2)}(x)$ has and only has one zero point within $x\in\mathbb{R}_+$ is analogous to that in the proof in Appendix~\ref{2nd domin proof}, and is therefore omitted here for brevity.

\section{Proof Details for Theorem~\ref{conv}}

\subsection{Proof of $f_{Y|W}(y|w)$}
\label{fYW}
To determine $f_{Y|W}(y|w)$, we first characterize $f_{Y|X,W}(y|x,w)$ before averaging it over the distribution of $X$, where $X$ is defined as a random variable to characterize the magnitude of the transmitted codeword. 

\begin{figure}
\centering
\includegraphics[width=0.5\textwidth]{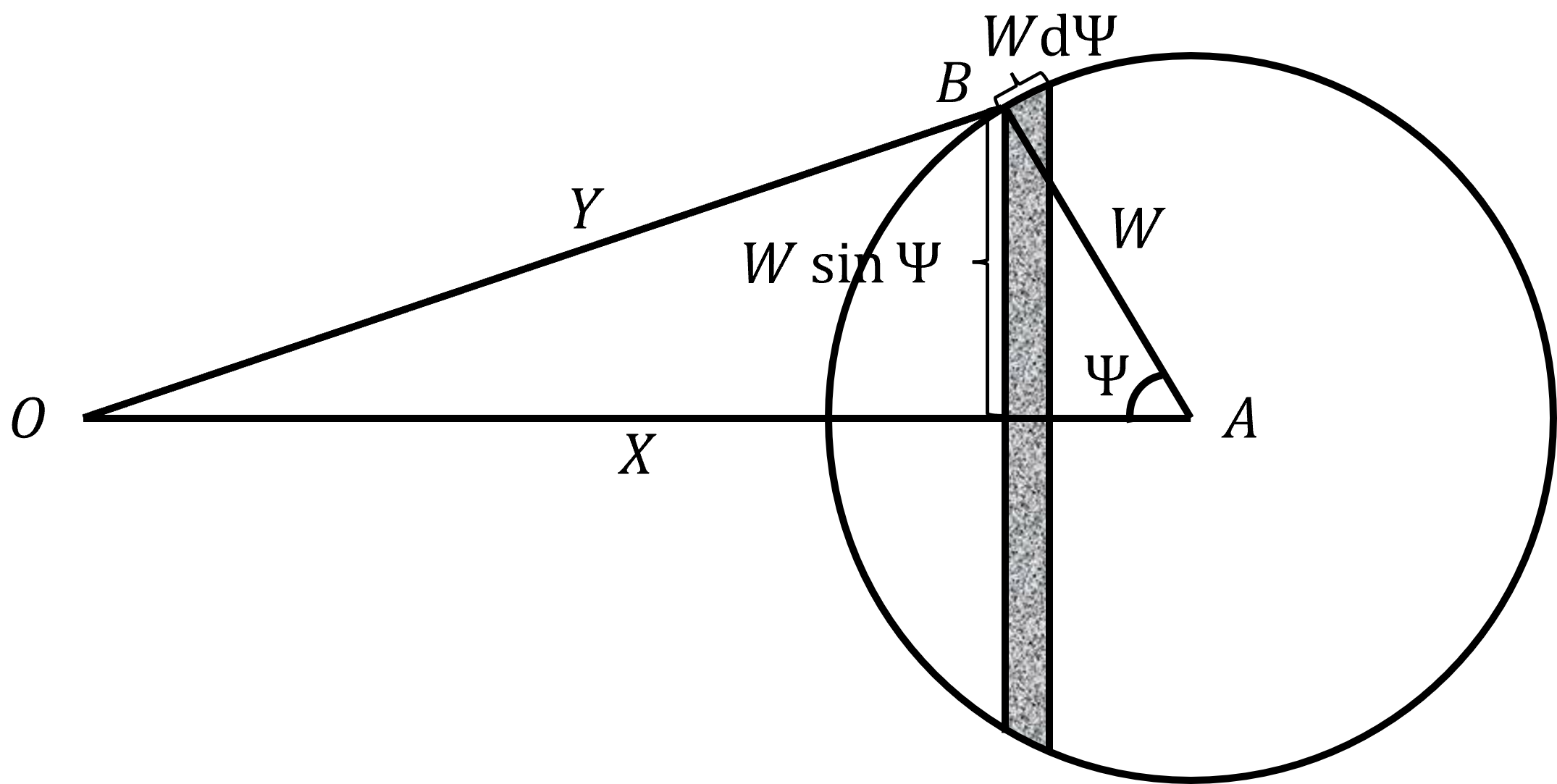}
\caption{Codewords with the same distance to the origin of the power sphere (projected surface view).}
\label{graph for lemma px}
\end{figure}

Since the noise is additive and isotropic, the received signal distributes uniformly on the surface of the codeword sphere. Fig.~\ref{graph for lemma px} illustrates this scenario, where $O$ is the origin of the power sphere, $A$ is the origin of the codeword, and $B$ is the location of the received signal. In addition, we have $OA=X$, $OB=Y$, $AB=W$, and $\angle OAB=\Psi$. All the points on the ring, i.e., the shaded area in Fig.~\ref{graph for lemma px}, share the same length of $OA$ and the same value of $\Psi$. Thus, 
\begin{align}
     f_{\Psi|X,W}(\psi,x,w)&=\frac{S_{\text{ring}}}{S_{\text{sphere}}\dd \Psi}=\frac{wS_{n-1}(w\sin\psi)}{S_{n}(w)}\\
     &=B\left(\frac{n-1}{2},\frac{1}{2}\right)^{-1}\sin^{n-2}\psi.
\end{align}
On substituting the PDF of $\Psi$ with $y^2=w^2+x^2-2wx\cos\psi$ and exploiting the transformation of random variables, we have
\begin{align}
f_{Y|X,W}(y|x,w)=f_{\Psi|Y,X,W}\left(\arccos \frac{w^2+x^2-y^2}{2wx}\right)\frac{1}{\sqrt{1-\frac{(w^2+x^2-y^2)^2}{4w^2x^2}}}\frac{y}{wx},\quad y\in[|x-w|,x+w].\label{1-1}
\end{align}

To determine $f_{Y|W}(y|w)$, we reiterate and simplify the expression for $f_{Y|X,W}(y|x,w)$ as
\begin{equation}
    f_{Y|X,W}(y|x,w)=f_\Psi(\psi)\frac{1}{\sin\psi}\frac{y}{wx},\,y\in[|x-w|,x+w],
\end{equation}
where
\begin{equation}
    \psi=\arccos\frac{w^2+x^2-y^2}{2wx}.
\end{equation}
Then, we have
    \begin{align}
        f_{Y|W}(y|w)&=\int_{0}^{\sqrt{h_jP_j}}f_X(x)f_{Y|X,W}(y|x,w)\,\dd x\\
        &=nB^{-1}\left(\frac{n-1}{2},\frac{1}{2}\right)(h_jP_j)^{-\frac{n}{2}}\frac{y}{w}\underbrace{\int_{y-w}^{\min\{y+w,\sqrt{h_jP_j}\}}x^{n-2}\left[1-\left(\frac{w^2+x^2-y^2}{2wx}\right)^2\right]^{\frac{n-3}{2}}\dd x}_{\triangleq A(y,w)}.\label{a5-1}
\end{align}
With some mathematical manipulation on $A(y,w)$, we have
\begin{align}
    A(y,w)&=(2w)^{3-n}\int_{y-w}^{\min\{y+w,\sqrt{h_jP_j}\}}x\left[4y^2w^2-\left(x^2-y^2-w^2\right)^2\right]^{\frac{n-3}{2}}\,\dd x \label{a5-3-1}\\
    &=\frac{1}{2}(2w)^{3-n}\int_{-2yw}^{\min\{2yw,h_jP_j-y^2-w^2\}}\left(4y^2w^2-p^2\right)^{\frac{n-3}{2}}\,\dd p\label{a5-3-2}\\
   &=y^{n-2}w\int_{-1}^{\min\{1,-m_T\}}\left(1-q^2\right)^{\frac{n-3}{2}}\,\dd q,\label{a5-3}
\end{align}
where \eqref{a5-3-1} and \eqref{a5-3-2} respectively employs $p=x^2-y^2-w^2$ and $q=\frac{p}{2yw}$, and \eqref{a5-3} uses $m_T=\frac{h_jP_j-y^2-w^2}{2yw}$. Further let $r=q^2$, we have
    \begin{equation}
\begin{aligned}
     A(y,w)=\frac{1}{2}y^{n-2}w\times\left\{\begin{array}{cl}
     \int_{\cos^2\beta}^{1}r^{-\frac{1}{2}}(1-r)^{\frac{n-3}{2}}\,\dd r,&m_T\leq0\\
     \int_{0}^{1}r^{-\frac{1}{2}}(1-r)^{\frac{n-3}{2}}\,\dd r+\int_{0}^{\cos^2\beta}r^{-\frac{1}{2}}(1-r)^{\frac{n-3}{2}}\,\dd r,&0<m_T<1\\
     2\int_{0}^{1}r^{-\frac{1}{2}}(1-r)^{\frac{n-3}{2}}\,\dd r,&m_T\geq1.
     \end{array}\right.\label{a5-2}
\end{aligned}
\end{equation}

Finally, by completing the integral and employing $\cos\beta=\frac{y^2+w^2-h_jP_j}{2yw}$, \eqref{2-4} is obtained.

\subsection{Proof of the Expressions for $\hat{\epsilon}_{j1|W}^{13}$ and $\hat{\epsilon}_{j1|W}^{23}$}
\label{ehat}

We first derive the expression for $\hat{\epsilon}_{j1|W}^{13}$. We start with
\begin{align}
    \tilde{\epsilon}_{j1|W}^{13}&\triangleq \frac{n}{4}(h_jP_j)^{-\frac{n}{2}}\int_{\sqrt{h_jP_j-w^2}}^{\sqrt{h_jP_j}+w} y^{n-1}I_{\sin^2\alpha}\left(\frac{n+1}{2},\frac{1}{2}\right)I_{\sin^2\beta}\left(\frac{n-11}{2},\frac{1}{2}\right)\,\dd y\\
    &\geq \frac{n}{n^2-1}B^{-1}\left(\frac{n+1}{2},\frac{1}{2}\right)B^{-1}\left(\frac{n-1}{2},\frac{1}{2}\right)(h_jP_j)^{-\frac{n}{2}}\int_{\sqrt{h_jP_j-w^2}}^{\sqrt{h_jP_j}+w} y^{n-1}\sin^{n+1}\alpha\sin^{n-1}\beta\,\dd y\label{ehat-1},
\end{align}
where \eqref{ehat-1} follows that
\begin{align}
    I_{z}\left(a,\frac{1}{2}\right)&=B^{-1}\left(a,\frac{1}{2}\right)\frac{z^a}{a}{_2F_1}\left(a,\frac{1}{2};a+1;z\right)\\
    &=B^{-1}\left(a,\frac{1}{2}\right)z^a\sum_{n=0}^{\infty}\frac{\left(\frac{1}{2}\right)_n}{n!(a+n)}z^n\\
    &\geq B^{-1}\left(a,\frac{1}{2}\right)z^a.
\end{align}
With sine theorem, we have
\begin{align}
    \frac{w}{\sin\alpha}=\frac{y}{\sin\beta}=2r,
\end{align}
where $r$ the radius of the circumcircle of a triangle with sides $x$, $y$ and $w$. Therefore, we have
\begin{align}
    y^{n-1}\sin^{n+1}\alpha\sin^{n-1}\beta&=y^{n-1}\left(\frac{w}{2R}\right)^{n+1}\left(\frac{y}{2R}\right)^{n-1}\\
    &=\frac{(2s)^{2n}}{y^{n+1}x^{n+1}w^{n-1}},
\end{align}
where $s=\frac{xyw}{4r}$ is the area of the circle with radius $r$. In addition, with Haron's formula \cite{Heron1896},
\begin{align}
    P(y)\triangleq\left(w+y+\sqrt{h_jP_j}\right)\left(-w+y+\sqrt{h_jP_j}\right)\left(w-y+\sqrt{h_jP_j}\right)\left(w+y-\sqrt{h_jP_j}\right),
\end{align}
we have
\begin{align}
    \tilde{\epsilon}_{j1|W}^{13} &\geq \frac{n}{n^2-1}B^{-1}\left(\frac{n+1}{2},\frac{1}{2}\right)B^{-1}\left(\frac{n-1}{2},\frac{1}{2}\right)(h_jP_j)^{-\frac{n}{2}}\int_{\sqrt{h_jP_j-w^2}}^{\sqrt{h_jP_j}+w} \frac{P(y)^n}{y^{n+1}}\,\dd y\\
    &\triangleq\hat{\epsilon}_{j1|W}^{13}.
\end{align}
The proof for the expression of $\hat{\epsilon}_{j1|W}^{23}$ follows analogous to that of $\hat{\epsilon}_{j1|W}^{13}$, and is therefore omitted for brevity.

\subsection{Proof of the Convexity of $\hat{\epsilon}_{j1|W}^{13}$ and $\hat{\epsilon}_{j1|W}^{23}$}
\label{ehat convex}

We first prove that $\hat{\epsilon}_{j1|W}^{13}$ is convex in $w$. By denoting $A\triangleq\frac{n}{n^2-1}B^{-1}\left(\frac{n+1}{2},\frac{1}{2}\right)B^{-1}\left(\frac{n-1}{2},\frac{1}{2}\right)\frac{(h_jP_j)^{n-\frac{1}{2}}}{4^n}$, $P(y)\triangleq-w^4+2(h_jP_j+y^2)w^2-(h_jP_j-w^2)^2$, $B(w)\triangleq\int_{\sqrt{h_jP_j-w^2}}^{\sqrt{h_jP_j}+w}C(y)\,\dd y$, and $C(y)=\frac{P(y)^n}{y^{n+1}}$, we have
\begin{align}
    \frac{\partial \hat{\epsilon}_{j1|W}^{13}}{\partial w}=A\left[(1-n)w^{-n}B(w)+w^{1-n}\frac{\partial B(w)}{\partial w}\right],
\end{align}
where
\begin{align}
    \frac{\partial B(w)}{\partial w}&=\int_{\sqrt{h_jP_j-w^2}}^{\sqrt{h_jP_j}+w}\frac{1}{x^{n+1}}\frac{\partial P(y)^n}{\partial w}\,\dd y +C\left(\sqrt{h_jP_j}+w\right)+\frac{w}{\sqrt{h_jP_j-w^2}}C\left(\sqrt{h_jP_j-w^2}\right)\\
    &\geq 0,
\end{align}
and
\begin{align}
    \frac{\partial P(y)^n}{\partial w}&=\frac{\partial P(y)^n}{\partial P(y)}\frac{\partial P(y)}{\partial w}\\
    &=4n wP(y)^{n-1}\left[h_jP_j+y^2-w^2\right]\\
    &\geq 0.
\end{align}
Then, we have
\begin{align}
    \frac{\partial^2 \hat{\epsilon}_{j1|W}^{13}}{\partial w^2}&=A\bigg\{(1-n)\left[-nw^{-n-1}B(w)+w^{-n}\frac{\partial B(w)}{\partial w}\right]\notag\\
    &\quad\quad\quad+\left[(1-n)w^{-n}\frac{\partial B(w)}{\partial w}+w^{1-n}\frac{\partial^2 B(w)}{\partial w^2}\right]\bigg\}\\
    &=A\left[w^{-n}(n-1)\left(nw^{-1}B(w)-2\frac{\partial B(w)}{\partial w}\right)+w^{1-n}\frac{\partial^2 B(w)}{\partial w^2}\right],
\end{align}
where
\begin{align}
    \frac{\partial^2B(w)}{\partial w^2}&=\int_{\sqrt{h_jP_j-w^2}}^{\sqrt{h_jP_j}+w}\frac{1}{y^{n+1}}\frac{\partial^2P(y)^n}{\partial w^2}\,\dd y+\frac{\partial C(\sqrt{h_jP_j})}{\partial w}+\frac{w}{\sqrt{h_jP_j-w^2}}\frac{\partial C(\sqrt{h_jP_j-w^2})}{\partial w}\notag\\
    &\quad+\frac{\partial C(\sqrt{h_jP_j}+w)}{\partial w}+\frac{\sqrt{h_jP_j-w^2}+\frac{w^2}{\sqrt{h_jP_j-w^2}}}{h_jP_j-w^2}C(\sqrt{h_jP_j-w^2})\notag\\
    &\quad+\frac{w}{\sqrt{h_jP_j-w^2}}\frac{\partial C(\sqrt{h_jP_j-w^2})}{\partial w}\\
    &=\int_{\sqrt{h_jP_j-w^2}}^{\sqrt{h_jP_j}+w}\frac{1}{y^{n+1}}\frac{\partial^2P(y)^n}{\partial w^2}\,\dd y+2\frac{\partial C(\sqrt{h_jP_j}+w)}{\partial w}\notag\\
    &\quad+\frac{2w}{\sqrt{h_jP_j-w^2}}\frac{\partial C(\sqrt{h_jP_j-w^2})}{\partial w}+\frac{h_jP_j}{(h_jP_j-w^2)^{\frac{3}{2}}}C(\sqrt{h_jP_j-w^2}).\label{pche-1}
\end{align}
In the following, we prove that all four terms in \eqref{pche-1} are nonnegative.
\subsubsection{$ \frac{\partial^2P(y)^n}{\partial w^2}$}
\begin{align}
    \frac{\partial^2P(y)^n}{\partial w^2}&=4n\left\{(-w^2+h_jP_j+y^2)P(y)^{n-1}+[-w^3+(h_jP_j+y^2)w]\frac{\partial P(y)^{n-1}}{\partial w}\right\}\\
    &=4n\left\{\right(-w^2+h_jP_j+y^2)P(y)^{n-1}\left[1+w^2(-w^2+h_jP_j+y^2)4n(n-1)P(y)^{n-2}\right]\}\\
    &\geq0.
\end{align}

\subsubsection{$\frac{\partial C(\sqrt{h_jP_j}+w)}{\partial w}$}
First, we calculate $C(\sqrt{h_jP_j}+w)$ as follows
\begin{align}
    C(\sqrt{h_jP_j}+w)&=\frac{P(\sqrt{h_jP_j}+w)^n}{(\sqrt{h_jP_j}+w)^{n+1}}\\
    &\begin{aligned}=\frac{1}{(\sqrt{h_jP_j}+w)^{n+1}}\Big[
    &-(\sqrt{h_jP_j}+w)^4+2(h_jP_j+w^2)(\sqrt{h_jP_j}+w)^2\notag\\
    &-(w^2-h_jP_j)^2\Big]\end{aligned}\\
    &=\frac{(2h_j^2P_j^2)^n}{(\sqrt{h_jP_j}+w)^{n+1}}.
\end{align}
Therefore, we have
\begin{align}
    \frac{\partial(\sqrt{h_jP_j}+w)}{\partial w}&=\frac{n+1}{2}(2h_j^2P_j^2)^n(\sqrt{h_jP_j}+w)^{-n-3}\\
    &\geq 0.
\end{align}
\subsubsection{$C(\sqrt{h_jP_j-w^2})$}
\begin{align}
    (\sqrt{h_jP_j-w^2})&=\frac{P(\sqrt{h_jP_j-w^2})^n}{(h_jP_j-w^2)^{\frac{n+1}{2}}}\\
    &\begin{aligned}=\frac{1}{(h_jP_j-w^2)^{\frac{n+1}{2}}}\Big[
    &-(h_jP_j-w^2)^2+2(h_jP_j+w^2)(h_jP_j-w^2)\notag\\
    &+(w^2-h_jP_j)^2\Big]\end{aligned}\\
    &=\frac{\left[2(h_jP_j+w^2)\right]^n}{(h_jP_j-w^2)^{\frac{n-1}{2}}}\\
    &\geq 0.
\end{align}
\subsubsection{$\frac{\partial C(\sqrt{h_jP_j-w^2})}{\partial w}$}
\begin{samepage}
\begin{align}
    \frac{\partial C(\sqrt{h_jP_j-w^2})}{\partial w}&=\frac{4w(h_jP_j-w^2)^{\frac{n-3}{2}}(h_jP_j+w^2)^{\frac{n-1}{2}}}{(h_jP_j-w^2)^{n-1}}\notag\\
    &\quad\times\left[n2^{n-1}(h_jP_j+w^2)^{\frac{n-1}{2}}(h_jP_j-w^2)+1)\right]\\
    &\geq 0.
\end{align}
\end{samepage}

The proof for the convexity of $\hat{\epsilon}_{j1|W}^{23}$ follows analogously to that of $\hat{\epsilon}_{j1|W}^{13}$. It is therefore omitted for brevity.
\end{appendices}

\bibliographystyle{IEEEtran}
\bibliography{reference}

\begin{thebibliography}{10}
\providecommand{\url}[1]{#1}
\csname url@samestyle\endcsname
\providecommand{\newblock}{\relax}
\providecommand{\bibinfo}[2]{#2}
\providecommand{\BIBentrySTDinterwordspacing}{\spaceskip=0pt\relax}
\providecommand{\BIBentryALTinterwordstretchfactor}{4}
\providecommand{\BIBentryALTinterwordspacing}{\spaceskip=\fontdimen2\font plus
\BIBentryALTinterwordstretchfactor\fontdimen3\font minus
  \fontdimen4\font\relax}
\providecommand{\BIBforeignlanguage}[2]{{%
\expandafter\ifx\csname l@#1\endcsname\relax
\typeout{** WARNING: IEEEtran.bst: No hyphenation pattern has been}%
\typeout{** loaded for the language `#1'. Using the pattern for}%
\typeout{** the default language instead.}%
\else
\language=\csname l@#1\endcsname
\fi
#2}}
\providecommand{\BIBdecl}{\relax}
\BIBdecl

\bibitem{you6G}
X.~You, W.~Xu, Y.~Huang, W.~Wang, F.~Zhu, and X.~Ji, ``Towards {6G} wireless
  communication networks: Vision, enabling technologies, and new paradigm
  shift,'' \emph{Sci. China Inf. Sci.}, vol.~64, no.~1, pp. 1--74, Nov. 2020.

\bibitem{10147111}
L.~Zhou and M.~Motani, ``Finite blocklength lossy source coding for discrete
  memoryless sources,'' \emph{Found. Trends Commun. Inf. Theory}, vol.~20,
  no.~3, pp. 157--389, 2023.

\bibitem{10024766}
W.~Xu, Z.~Yang, D.~W.~K. Ng, M.~Levorato, Y.~C. Eldar, and M.~Debbah, ``Edge
  learning for {B5G} networks with distributed signal processing: Semantic
  communication, edge computing, and wireless sensing,'' \emph{IEEE J. Sel.
  Topics in Signal Process.}, vol.~17, no.~1, pp. 9--39, Jan. 2023.

\bibitem{8187464}
V.~Y.~F. Tan and A.~Roth, ``Asymptotic estimates in information theory with
  non-vanishing error probabilities,'' \emph{Found. Trends Commun. Inf.
  Theory}, vol.~11, no. 1-2, pp. 1--184, 2014.

\bibitem{PPV}
Y.~Polyanskiy, H.~V. Poor, and S.~Verd\'u, ``Channel coding rate in the finite
  blocklength regime,'' \emph{IEEE Trans. Inf. Theory}, vol.~56, no.~5, pp.
  2307--2359, May 2010.

\bibitem{Tan}
V.~Y.~F. Tan and M.~Tomamichel, ``The third-order term in the normal
  approximation for the {AWGN} channel,'' \emph{IEEE Trans. Inf. Theory},
  vol.~61, no.~5, pp. 2430--2438, May 2015.

\bibitem{YKE}
R.~C. Yavas, V.~Kostina, and M.~Effros, ``Third-order analysis of channel
  coding in the small-to-moderate deviations regime,'' \emph{IEEE Trans. Inf.
  Theory}, vol.~70, no.~9, pp. 6139--6170, Sep. 2024.

\bibitem{SP59}
C.~E. Shannon, ``Probability of error for optimal codes in a {Gaussian}
  channel,'' \emph{The Bell Syst. Tech. J.}, vol.~38, no.~3, pp. 611--656, May
  1959.

\bibitem{SP67}
R.~Gallager, ``A simple derivation of the coding theorem and some
  applications,'' \emph{IEEE Trans. Inf. Theory}, vol.~11, no.~1, pp. 3--18,
  Jan. 1965.

\bibitem{1967}
\BIBentryALTinterwordspacing
C.~Shannon, R.~Gallager, and E.~Berlekamp, ``Lower bounds to error probability
  for coding on discrete memoryless channels. {I},'' \emph{Inf. Cont.},
  vol.~10, no.~1, pp. 65--103, Jan. 1967. [Online]. Available:
  \url{https://www.sciencedirect.com/science/article/pii/S0019995867900526}
\BIBentrySTDinterwordspacing

\bibitem{ML}
E.~MolavianJazi and J.~N. Laneman, ``A second-order achievable rate region for
  {Gaussian} multi-access channels via a central limit theorem for functions,''
  \emph{IEEE Trans. Inf. Theory}, vol.~61, no.~12, pp. 6719--6733, Oct. 2015.

\bibitem{scarlettMAC}
J.~Scarlett, A.~Martinez, and A.~Guillén~i Fàbregas, ``Second-order rate
  region of constant-composition codes for the multiple-access channel,''
  \emph{IEEE Trans. Inf. Theory}, vol.~61, no.~1, pp. 157--172, Nov. 2015.

\bibitem{YKEMAC}
R.~C. Yavas, V.~Kostina, and M.~Effros, ``{Gaussian} multiple and random access
  channels: Finite-blocklength analysis,'' \emph{IEEE Trans. Inf. Theory},
  vol.~67, no.~11, pp. 6983--7009, Sep. 2021.

\bibitem{zhoulin}
Y.~Wang, L.~Bai, Z.~Wu, and L.~Zhou, ``Achievable second-order asymptotics for
  {MAC} and {RAC} with additive non-{Gaussian} noise,'' \emph{IEEE Trans. Inf.
  Theory}, vol.~71, no.~7, pp. 4953--4975, May 2025.

\bibitem{Tuninetti2023}
D.~Tuninetti, P.~Sheldon, B.~Smida, and N.~Devroye, ``On second order rate
  regions for the static scalar {Gaussian} broadcast channel,'' \emph{IEEE J.
  Sel. Areas Commun.}, vol.~41, no.~7, pp. 1982--1999, Jul. 2023.

\bibitem{UnsalGorce2026}
A.~{\"U}nsal and J.-M. Gorce, ``Achievability bounds of coding with finite
  blocklength for gaussian broadcast channels,'' 2026.

\bibitem{Tanconv}
S.~L. Fong and V.~Y.~F. Tan, ``A proof of the strong converse theorem for
  {Gaussian} multiple access channels,'' \emph{IEEE Trans. Inf. Theory},
  vol.~62, no.~8, pp. 4376--4394, May 2016.

\bibitem{Kosutconv}
O.~Kosut, ``A second-order converse bound for the multiple-access channel via
  {Wringing} dependence,'' \emph{IEEE Trans. Inf. Theory}, vol.~68, no.~6, pp.
  3552--3584, Feb. 2022.

\bibitem{TIT}
K.~Bao, W.~Xu, X.~You, and H.~V. Poor, ``New sphere-packing bounds for
  finite-blocklength coding over additive noise channels,'' \emph{IEEE Trans.
  Inf. Theory}, vol.~72, no.~7, pp. 4468--4490, May 2026.

\bibitem{10.5555/39091}
J.~H. Conway, N.~J.~A. Sloane, and E.~Bannai, \emph{Sphere-packings, lattices,
  and groups}.\hskip 1em plus 0.5em minus 0.4em\relax Berlin, Heidelberg:
  Springer-Verlag, 1987.

\bibitem{KN1}
E.~Yves, R.~E.~M., and N.~J.~A. Sloane, ``On kissing numbers in dimensions 32
  to 128,'' \emph{Electronic J. Combinatorics}, vol.~22, no.~1, pp. 1--5, Apr.
  1998.

\bibitem{7300429}
E.~MolavianJazi and J.~N. Laneman, ``A second-order achievable rate region for
  {Gaussian} multi-access channels via a central limit theorem for functions,''
  \emph{IEEE Trans. Inf. Theory}, vol.~61, no.~12, pp. 6719--6733, Oct. 2015.

\bibitem{11006123}
Y.~Wang, L.~Bai, Z.~Wu, and L.~Zhou, ``Achievable second-order asymptotics for
  {MAC} and {RAC} with additive non-{Gaussian} noise,'' \emph{IEEE Trans. Inf.
  Theory}, vol.~71, no.~7, pp. 4953--4975, May 2025.

\bibitem{9535162}
R.~C. Yavas, V.~Kostina, and M.~Effros, ``{Gaussian} multiple and random access
  channels: Finite-blocklength analysis,'' \emph{IEEE Trans. Inf. Theory},
  vol.~67, no.~11, pp. 6983--7009, Sep. 2021.

\bibitem{Mross2024}
M.~A. Mross, P.-H. Lin, and E.~A. Jorswieck, ``{Gaussian} broadcast channels
  with heterogeneous finite blocklength constraints: Inner and outer bounds,''
  \emph{IEEE Trans. Commun.}, vol.~72, no.~5, pp. 2731--2745, May 2024.

\bibitem{Scarlett}
J.~Scarlett, V.~Y.~F. Tan, and G.~Durisi, ``The dispersion of nearest-neighbor
  decoding for additive non-{Gaussian} channels,'' \emph{IEEE Trans. Inf.
  Theory}, vol.~63, no.~1, pp. 81--92, Jan. 2017.

\bibitem{StochasticOrders}
M.~Shaked and J.~G. Shanthikumar, \emph{Stochastic Orders}.\hskip 1em plus
  0.5em minus 0.4em\relax New York, NY, USA: Springer, 2007.

\bibitem{Heron1896}
H.~Alexandria, \emph{Metrica}.\hskip 1em plus 0.5em minus 0.4em\relax
  Cambridge, U.K.: Cambridge University Press, 1896.

\end{thebibliography}

\end{document}